\documentclass[%
    reprint,
    superscriptaddress,
    letterpaper,
    amsmath,
    amssymb,
    aps,
    prx,
    floatfix,
]{revtex4-2}

\usepackage{graphicx}
\usepackage[usenames, dvipsnames, x11names]{xcolor}
\usepackage[colorlinks=true,linkcolor=blue,citecolor=blue]{hyperref}

\usepackage[capitalise]{cleveref}
\crefname{methodsfigure}{Extended Data Figure}{Extended Data Figures}

\usepackage{siunitx}
\usepackage{bm}
\usepackage{mathtools}
\usepackage{braket}

\usepackage{setspace}
\usepackage[
    textwidth=0.6in,
    textsize=footnotesize,
    color=red!25,
    colorinlistoftodos,
]{todonotes}

\usepackage{soul}
\setstcolor{red}

\begin{document}
\title{Parametrically controlled chiral interface for superconducting quantum devices}

\author{Xi Cao}
\email{xicao@illinois.edu}
\author{Abdullah Irfan}
\author{Michael Mollenhauer}
\author{Kaushik Singirikonda}
\affiliation{Department of Physics, University of Illinois Urbana-Champaign, Urbana, IL 61801, USA}
\author{Wolfgang Pfaff}
\email{wpfaff@illinois.edu}
\affiliation{Department of Physics, University of Illinois Urbana-Champaign, Urbana, IL 61801, USA}
\affiliation{Materials Research Laboratory, University of Illinois Urbana-Champaign, Urbana, IL 61801, USA}

\begin{abstract}
Nonreciprocal microwave routing plays a crucial role for measuring quantum circuits, and allows for realizing cascaded quantum systems for generating and stabilizing entanglement between non-interacting qubits.
The most commonly used tools for implementing directionality are ferrite-based circulators.
These devices are versatile, but suffer from excess loss, a large footprint, and fixed directionality.
For utilizing nonreciprocity in scalable quantum circuits it is desirable to develop  efficient integration of low-loss and in-situ controllable directional elements.
Here, we report the design and experimental realization of a minimal controllable directional interface that can be directly coupled to superconducting qubits.
In the presented device, nonreciprocity is realized through a combination of interference and phase-controlled parametric pumping.
We have achieved a maximum directionality of around 30\,dB, and the performance of the device is predicted quantitatively from independent calibration measurements.
Using the excellent agreement of model and experiment, we predict that the circuit will be useable as a chiral qubit interface with inefficiencies at the one-percent level or below.
Our work offers a promising route for realizing high-fidelity signal routing and entanglement generation in all-to-all connected microwave quantum networks, and provides a path for isolator-free qubit readout schemes.
\end{abstract}
 
\maketitle

\section{Introduction}
Nonreciprocal signal routing is an essential  ingredient for the practical operation of quantum devices as well as for the observation of a range of fundamental quantum phenomena.
For one, nonreciprocity enables efficient yet noise-protected signal detection, which is a foundation of quantum-limited measurement in cryogenic quantum circuits~\cite{boissonneault_dispersive_2009, abdo_active_2019, lecocq_Efficient_2021, rosenthal_Efficient_2021, abdo_High_2021}. 
The ability to perform high-fidelity qubit readout with minimal back-action, in particular, will be crucial for fault-tolerant quantum computation~\cite{divincenzo_faulttolerant_2009, fowler_surface_2012, devoret_superconducting_2013}.
Single-shot readout with fidelities above $99 \%$ has been demonstrated for superconducting qubits using readout chains with commercial circulators~\cite{walter_rapid_2017, sunada_fast_2022}.
On the other hand, directional propagation of photons allows the realization of cascaded quantum systems~\cite{gardiner_input_1985, gardiner_driving_1993, carmichael_quantum_1993, lodahl_chiral_2017} that enable the generation and autonomous stabilization of remote entanglement~\cite{cirac_quantum_1997, stannigel_drivendissipative_2012}, providing a route for modular scaling of quantum circuits.

While conventional microwave circulators have allowed the realization of prototypical cascaded systems of superconducting qubits~\cite{axline_ondemand_2018, campagne-ibarcq_deterministic_2018, kurpiers_deterministic_2018}, achieving practical utility will require efficient directional interfaces that are directly integrated with qubits.
The central functionality for such an interface is the controllable directional coupling of a superconducting circuit mode that stores or processes quantum information to a transmission line.
Tunable directionality can be used for on-demand quantum signal routing, enabling for interconnects for highly connected networks of quantum devices~\cite{gheeraert_programmable_2020}. 
Beyond this basic functionality, we require efficiency and linearity.
An efficient directional interface introduces no loss channels on the quantum mode, and can faithfully emit (absorb) quantum information into (from) propagating states.
High efficiency and low impact on circuit coherence is required to improve the fidelity of quantum state transmission which is currently limited by the loss from the stand-alone circulators~\cite{axline_ondemand_2018, campagne-ibarcq_deterministic_2018, kurpiers_deterministic_2018}.
Finally, linearity allows for transferring arbitrary quantum states ranging from single-photon encodings to error correctable bosonic multi-photon states.

\begin{figure*}[t]
\centering
\includegraphics{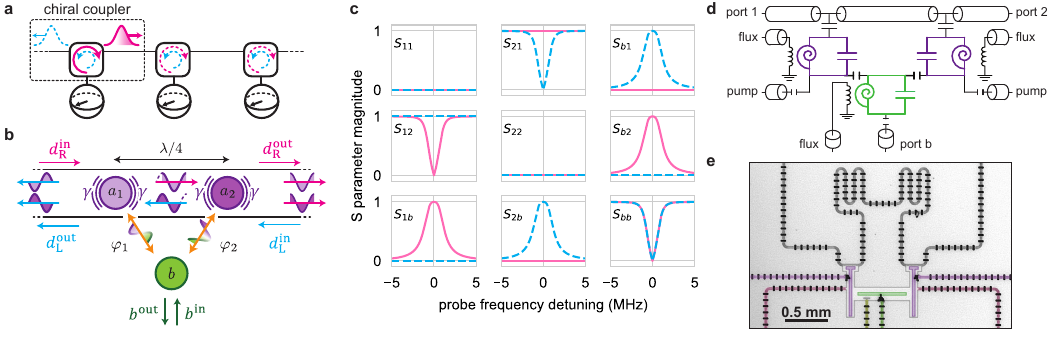}
	\caption{
    \textbf{Chiral coupler concept and design.}
    \textbf{a,} Our goal is a controllable circulator that may be seamlessly integrated with superconducting qubits, independent of encoding. 
    It may be used to achieve all-to-all connected quantum state transfer in modular quantum processors.
    \textbf{b,} Minimal model for a parametrically controlled circulator.
    Modes $a_{1,2}$ have frequency $\omega_0$, and couple to left- and right-propagating modes with rate $\gamma$.
    Combining phase-controlled mode-conversion between $b$ and $a_{1,2}$ with propagation delay of emitted/absorbed photons results in tuneable circulation through constructive/destructive interference at the right/left ports.
    \textbf{c,} Calculated S-matrix. 
    Parameters used here: $\gamma = \gamma_b = 2\pi\times\qty{1}{\mega\hertz}$, $g_1 = g_2 = 2\pi \times \qty{0.5}{\mega\hertz}$; $\varphi_1 = 0,~\varphi_2 = \pi/2$ (pink), $\varphi_1 = 0,~\varphi_2 = -\pi/2$ (blue).
    \textbf{d,} Lumped-element equivalent circuit. 
    Mode $b$ is frequency-tuneable in this implementation.
    Each mode is capacitively coupled to an external port for measuring S-parameters.
    Two weakly coupled ($Q_c > 10^6$) ports are attached to the $a_1$ and $a_2$ modes for parametric pumping.
    Each SNAIL is flux-tunable.
    \textbf{e,} False-colored optical image of the central portion of the device presented in this work.
    }
\label{fig:overview}
\end{figure*}

There are multiple possible avenues for realizing integrated nonreciprocity.
One strategy is to introduce an appropriate material into the circuit and bias it with an external magnetic field to break symmetry~\cite{mahoney_onchip_2017, wang_lowloss_2021, wang_dispersive_2024, owens_chiral_2022}. 
The need for applying an external field to a part of the circuit is, however, inherently at odds with preserving high coherence of superconducting qubits.
Alternatively, it is possible to realize directionality using synthetic fields in Josephson circuits~\cite{koch_timereversalsymmetry_2010, metelmann_nonreciprocal_2015, kerckhoff_onchip_2015, kamal_minimal_2017, lecocq_nonreciprocal_2017, muller_passive_2018, naghilooBroadbandMicrowaveIsolation2021, zhuang_superconducting_2023}.
This approach has the unique advantage of immediate compatibility with superconducting qubits, which is appealing for realizing efficient, integrated devices.
A range of stand-alone and general-purpose microwave circulators and isolators have been demonstrated in recent years~\cite{sliwa_reconfigurable_2015, chapman_widely_2017, abdo_active_2019, ranzani_circulators_2019, huang_design_2022, beck_wideband_2023, kwende_josephson_2023, navarathna_passive_2023}. 
Moreover, superconducting qubits that couple nonreciprocally to propagating microwaves in waveguide quantum electrodynamics (wQED) have been proposed and experimentally realized \cite{gheeraert_programmable_2020, guimond_unidirectional_2020, zhang_chargenoise_2021, redchenko_tunable_2023, kannan_ondemand_2023, joshi_resonance_2023, yen_allpass_2024b}. 
Operating on the principle of so-called `giant atoms' \cite{kannan_waveguide_2020}, hallmarks of chiral wQED such as directional emission \cite{kannan_ondemand_2023} and scattering \cite{joshi_resonance_2023} were observed.

Here, we propose and experimentally demonstrate a general chiral interface --- a `chiral coupler' --- that is suited for integration with (arbitrary) superconducting quantum devices and will allow for high efficient quantum state emission and absorption. 
We approach the following discussion predominantly from a quantum network perspective, where the goal is to route quantum states between qubits using directional emission and absorption (\cref{fig:overview}a); our circuit is, however, well-suited also for integrated noise isolation \cite{chapman_widely_2017, abdo_active_2019, beck_wideband_2023, kwende_josephson_2023, navarathna_passive_2023}.

The central considerations of our design are as follows:
We target a device with three modes (\cref{fig:overview}b):
Two modes couple to a common transmission line at two physically separate points and operate at the same communication frequency.
The third mode is tunably coupled to the transmission line via parametric mode-conversion with the other two modes, similar to the idea of the giant atom in waveguide QED~\cite{friskkockum_designing_2014, kockum_decoherencefree_2018, vadiraj_engineering_2021, kannan_waveguide_2020, wang_chiral_2022, yin_generation_2023}. 
This mode may be the coherent quantum mode of interest (e.g., a superconducting qubit or bosonic memory element), or coherently couple to such a mode.

These considerations directly address our key criteria for practical utility.
The direct integration of the chiral interface with the qubit mode is important for achieving high efficiency, as it minimizes the number of couplings and connections.
We emphasize that our design only uses the minimal number of modes to achieve nonreciprocity~\cite{clerk_introduction_2022}.
In this way, we can analyze the device using a simple yet predictive model, and minimize the number of additional decoherence channels introduced to the quantum circuit.
Our envisioned use also motivates that we do not require a large bandwidth, but merely some degree of (static) frequency tuning capability.
Finally, the circuit supports isolation and circulation of arbitrary quantum states; 
directional emission (absorption) from (to) qubits have been demonstrated recently using transmon-based couplers~\cite{kannan_ondemand_2023, almanakly_Deterministic_2024}. 
Our circuit, in contrast, is linear and will allow for transmitting bosonic quantum error correction codewords~\cite{ofekExtendingLifetimeQuantum2016, sivakRealtimeQuantumError2023}.

We note that linearity also allows the circuit to be used as a replacement for conventional circulators employed in established qubit measurement techniques using coherent states.
In an experimental realization of the proposed circuit we have achieved high directionality ($\sim 30$~dB), determined through measurements of isolation and circulation.
As result of its minimal design, we can model the device performance in quantitative agreement with a few independently calibrated circuit parameters.
With the circuit parameters chosen in the device measured here, directional emission and absorption of arbitrary quantum states in a network could be achieved with an efficiency of 85\%.
Modest adjustment of circuit parameters will bring this efficiency to the 99\%-level.
The circuit presented thus provides a promising, flexible chiral interface for superconducting qubit devices for the realization of low-loss cascaded networks and for integrated noise-isolation.

\section{Model}
\label{sec:model}
Our system is composed of three modes, in a similar spirit to recently investigated giant atoms in waveguide QED, illustrated in \cref{fig:overview}b. \cite{friskkockum_designing_2014, kockum_decoherencefree_2018, vadiraj_engineering_2021, kannan_waveguide_2020, wang_chiral_2022, yin_generation_2023}, 
We realize directionality by phase-controlled interference between different emission paths. 
Two modes $a_{1,2}$ with identical frequency couple to a common waveguide with strength $2 \gamma_{1, 2}$.
They are spatially separated along the waveguide by a distance $\Delta x = \lambda/4$, where $\lambda$ is the wavelength of the emitted radiation. 
Mode $b$ couples to both $a_{1, 2}$ via parametrically controllable mode conversion, with rate $g_{1, 2}$ and phase $\varphi_{1, 2}$.
A photon from $b$ may enter the waveguide via two paths: $b \rightarrow a_1$ and $b \rightarrow a_2$. 
This setup enables phase control of the emission paths through external drives, and thus controllable interference between the paths, which is the origin of nonreciprocity here.
The only requirement on the modes in the circuit is that phase-controlled frequency conversion can be established between them.
There is no restriction on the nature of the $b$ mode, and it could serve as the quantum mode of interest or directly couple to it. 
The chiral coupler could thus act as a general-purpose, narrow-band circulator that is directly hybridized with a coherent quantum mode.

We model the system as a three-port device, where $d^{\text{in(out)}}_\text{L}$,  $d^{\text{in(out)}}_\text{R}$, $b^{\text{in(out)}}$ are the input (output) field operators at each port (\cref{fig:overview}d).
The equations of motions are
\begin{align}
    \dot{a}_1 = &-i\omega_0 a_1 -i g_{1} e^{i (\omega_\text{p} t + \varphi_{1})} b - \gamma a_1 \nonumber \\
    &- i \gamma a_2 - i g_\text{c} a_2 - \sqrt{\gamma}  d_\text{L}^\text{in} - \sqrt{\gamma}  d_\text{R}^\text{in}, \label{Eq:a equation} \\
    \dot{b} = &-i\omega_b b -i g_{1} e^{-i (\omega_p t + \varphi_{1})} a_1  \nonumber \\ 
    &-i g_{2} e^{-i(\omega_p t +  \varphi_{2})} a_2 - \frac{\gamma_b}{2} b - \sqrt{\gamma_b} b^{\text{in}}, \label{Eq:b equation} \\
    \dot{a}_2 = &-i\omega_0 a_2 -i g_{2} e^{i(\omega_p t +  \varphi_{2})} b - \gamma a_2 \nonumber \\ 
    &- i \gamma a_1 - i g_\text{c} a_1 - \sqrt{\gamma} e^{-i \frac{\pi}{2}} d_\text{L}^{\text{in}} - \sqrt{\gamma} e^{i \frac{\pi}{2}} d_\text{R}^{\text{in}} \label{Eq:c equation}.
\end{align}
We assume for now that both $a_1$ and $a_2$ couple to the waveguide equally, i.e., $\gamma_1 = \gamma_2 = \gamma$, and $\gamma_b$ is the coupling strength between $b$ and its port. 
We first focus on the ideal case with no internal damping, i.e., $\gamma_{\text{i}, 1} = \gamma_{\text{i}, 2} = \gamma_{\text{i}, b} = 0$.
$\omega_0$ and $\omega_b$ are the mode frequencies of $a_{1, 2}$ and $b$ respectively.
At the operation point, both mode $a_1$ and $a_2$ have the same frequency, i.e., $\omega_0 = \omega_1 = \omega_2$.
The terms $i\gamma a_2$ in \cref{Eq:a equation} and $i \gamma a_1$ in \cref{Eq:c equation} arise from a waveguide-mediated interaction between $a_1$ and $a_2$.
Crucially, this interaction must be fully eliminated for the chiral coupler to operate as a circulator \cite{gheeraert_programmable_2020} (\cref{fig: residual coupling}).
We can eliminate this interaction by cancelling it with a bus-mediated coupling~\cite{majer_coupling_2007, chen_qubit_2014, lu_universal_2017, yan_tunable_2018}:
The terms $i g_\text{c} a_2$ in \cref{Eq:a equation} and $i g_\text{c} a_1$ in \cref{Eq:c equation} are the result of this coupling, arising from a static coupling between $a_{1,2}$ to $b$, with
\begin{equation}
    g_\text c = \frac{g_{a_1 b} g_{a_2 b}}{\Delta}.
\end{equation}
Here, $g_{a_{1(2)}b}$ is the static coupling between the undressed $a_{1(2)}$ and $b$ modes, and $\Delta = \omega_b - \omega_{1(2)}$.
$g_\text c$ (and in particular, its sign) can be controlled through the detuning between the $a_{1,2}$ and $b$ (see \cref{app:coupling cancellation} for details).

If the cancellation condition is met, i.e., $g_\text{c} = -\gamma$, circulation can be seen directly from the S-matrix, which is plotted as function of probe frequency detuning at each port in \cref{fig:overview}c.
We have derived the S-parameters by combining the equations of motion with the input-output relations, which are given by
\begin{align}
    d_\text{L}^{\text{out}} &=  d_\text{L}^{\text{in}} + \sqrt{\gamma} (a_1 + e^{i \frac{\pi}{2}} a_2 ),\\
    d_\text{R}^\text{out} &=  d_\text{R}^{\text{in}} + \sqrt{\gamma} (a_1 + e^{-i \frac{\pi}{2}} a_2 ),\\
    b^{\text{out}} &= b^{\text{in}} + \sqrt{\gamma_\text{b}} b.
    \label{Eq:input-output relation}
\end{align}
Full expressions for the S-matrix as well as details of the derivation are given in \cref{app:S-matrix}. 
Of particular importance is the waveguide transmission, $S_{21}$.
For perfect cancellation and without internal damping and parametric pumping, it is given by
\begin{equation}
    S_{21}(\delta) = \frac{\delta^2 + \gamma^2}{(\delta + i \gamma)^2},
\label{Eq: S21 cancellation}
\end{equation}
where $\delta$ is the detuning of the probe signal from the mode frequency.
In the following section, we will show in detail that the cancellation condition can be calibrated by measuring this transmission.


\section{Circuit realization and characterization}
\label{sec:circuit}
Our model can be realized with the superconducting circuit shown in \cref{fig:overview}d,e.
It consists of three modes, each containing a \emph{Superconducting Nonlinear Asymmetric Inductive eLement} (SNAIL)~\cite{frattini_3wave_2017} and a common transmission line. 
We have chosen the SNAIL elements for the following reasons:
SNAILs provide a third-order nonlinearity that enables parametric frequency conversion between the modes.
Their frequency-dependence on external flux can be used to tune mode frequencies to the operation point.
Importantly, charge-coupled SNAILs haven been shown to be highly coherence-preserving when used as couplers between bosonic modes~\cite{chapman_highonoffratio_2023}.
For this reason, we anticipate them to be a good choice for achieving high efficiencies.

This minimal circuit can be understood quantitatively with a small set of parameters.
In the following, we first describe how the device parameters shown in \cref{tab:params} are extracted from experiment.
Further below, we show that these parameters are sufficient to predict the circulator performance of the device in quantitative agreement with experiment.

\begin{table}[htb]
    \caption{\label{tab:params}
    Circuit parameters entering the model. Values are the ones at the operation point, and are extracted from spectroscopy data.
    }
    \begin{ruledtabular}
    \begin{tabular}{rll}
    \textrm{Qty.} & \textrm{Value} & \textrm{Description}\\
    \colrule
    $\omega_1/2\pi$ & \qty{4.875}{\giga\hertz} & resonant frequency of $a_1$\\
    $\omega_2/2\pi$ & \qty{4.875}{\giga\hertz} & resonant frequency of $a_2$\\
    $\omega_b/2\pi$ & \qty{6.270}{\giga\hertz} & resonant frequency of $b$\\
    $\gamma_{\text{i},1}/2\pi$ & \qty{215}{\kilo\hertz} & internal damping of $a_1$\\
    2$\gamma_{1}/2\pi$ & \qty{1.46}{\mega\hertz} & waveguide coupling of $a_1$\\
    $\gamma_{\text{i},2}/2\pi$ & \qty{294}{\kilo\hertz} & internal damping of $a_2$\\
    2$\gamma_{2}/2\pi$ & \qty{1.43}{\mega\hertz} & waveguide coupling of $a_2$\\
    $\gamma_{\text{i},b}/2\pi$ & \qty{588}{\kilo\hertz} & internal damping of $b$\\
    $\gamma_{b}/2\pi$ & \qty{2.51}{\mega\hertz} & waveguide coupling of $b$\\
    $g_\text c/\gamma$ & \qty{1.04}{ } & coupling cancellation ratio \\
    $g_{1(2)} /2\pi$ & \qty{0.70}{\mega\hertz} & parametric coupling strength \\
    \end{tabular}
    \end{ruledtabular}
\end{table}

\begin{figure}
\centering
\includegraphics{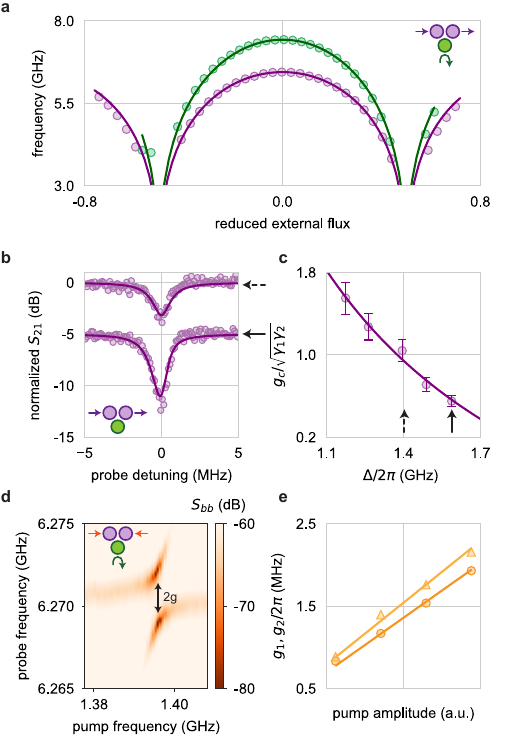}
\caption{\textbf{Mode characterization.}
    \textbf{a,} $a_1$ (purple) and $b$ (green) resonant frequencies vs.\ respective external flux, extracted from spectroscopy.
    \textbf{b,} Normalized $S_{21}$ at two different values of $\Delta$, 
    $g_\text{c}/\sqrt{\gamma_1 \gamma_2}=1.04$, with $g_\text{c}/2\pi = 0.75$~MHz (top curve, dashed-line arrow) and $g_\text{c}/\sqrt{\gamma_1 \gamma_2}=0.55$, with $g_\text{c}/2\pi = 0.40$~MHz (bottom curve, solid-line arrow).
    Data were normalized by dividing out the background. 
    Lower curve is shifted by $-5$\,dB for clarity. 
    \textbf{c,} Ratio between the bus-mediated and waveguide-mediated coupling as a function of detuning between $a_{1,2}$ and $b$.
    $g_\text{c}$ is obtained by a fit of $S_{21}$ to a full model.
    Error bar indicates a 10\% variation of the fit parameter, estimated from the least-squares error. 
    \textbf{d,} Normal-mode splitting between $a_1$ and $b$ induced by parametric pumping.
    $b$ is measured in reflection while frequency conversion pumps are applied. 
    \textbf{e,} Calibration of the coupling strength $g_1$ (light orange triangles) and $g_2$ (dark orange circles) as a function of pump voltage.
    Lines are linear fits.
}
\label{fig:basic calibration}
\end{figure}

We have performed basic mode characterization at 10~mK prior to applying parametric pumps using a vector network analyzer (VNA).
In \cref{fig:basic calibration}a we show the mode frequencies as function of external flux.
We have obtained resonant frequencies, external coupling and internal damping by standard fitting of the VNA trace for each flux bias point.
At the targeted operation point, $\omega_1 = \omega_2 \equiv \omega_0$ and $\Delta \equiv \omega_b - \omega_0$ is chosen such that $g_\text c = -\gamma$.
The effect of the waveguide-mediated coupling is manifested in the waveguide transmission $S_{21}$, \cref{Eq: S21 cancellation}; this allows us to use waveguide transmission to find the cancellation condition.
Minimal waveguide-induced coupling corresponds to highest transmission at $\omega_0$.
To find the best operation point, we have thus measured $S_{21}$ as function of $\omega_b$, controlled by external flux, while keeping $\omega_0$ fixed.
$g_\text{c}$ was inferred from a fit to $S_{21}(\delta)$ (see \cref{Eq: S21 eq nonideal}).
Two representative VNA traces and corresponding fits are shown in \cref{fig:basic calibration}b.
We note that ideally one can expect unit transmission for perfect cancellation;
finite internal damping or dephasing, however, result in residual insertion loss at $\omega_0$.
We note that in practice we cannot expect an exact matching of external coupling rates ($\gamma_1 \neq \gamma_2$), and the cancellation condition slightly changes: $g_\text{c} = -\sqrt{\gamma_1 \gamma_2}$ (\cref{app:S-matrix}). 
In \cref{fig:basic calibration}c, we show the inferred cancellation ratio $g_\text c/\sqrt{\gamma_1 \gamma_2}$ as a function of $\Delta$, following the expected $1/\Delta$ dependence.
The operation point in what follows is at $\Delta/2\pi = \qty{1.395}{\giga\hertz}$, where we found the best cancellation.

Next, we turn to the realization of parametric frequency conversion between modes $a_{1(2)}$ and $b$.
By applying external pump tones at frequency $\Delta$ we can enable effective `beam splitter' Hamiltonians~\cite{aumentado_superconducting_2020, chapman_highonoffratio_2023, zhou_realizing_2023},
\begin{equation}
    H_{g_{1(2)}} = g_{1(2)} e^{i\varphi_{1(2)}} a_{1(2)}^{\dagger} b + \text{h.c.},
\end{equation}
where the effective coupling strengths $g_{1(2)}$ and phases $\varphi_{1(2)}$ are controlled by the pumps.
To determine $g_{1(2)}$, we probe the reflection off $b$ while applying a pump with frequency $\omega_{\text{p}_{1(2)}} \sim \Delta$; a representative response signal is shown in \cref{fig:basic calibration}d. 
At the resonance condition, $\omega_{\text{p}_{1(2)}} = \omega_b - \omega_{a_{1(2)}}$, we observe an anti-crossing with a mode separation of $2 g_{1(2)}$. 
Coupling strengths $g_{1(2)}$ determined in this way are shown as a function of pump amplitude in \cref{fig:basic calibration}e. 
Coupling strength increases linearly with pump amplitude, in agreement with our expectation for a parametrically driven three-wave mixing element~\cite{chapman_highonoffratio_2023, zhou_realizing_2023}.

\section{Isolation and Circulation}
\label{sec:isolation-and-gyration}

\begin{figure}[bt]
\centering
\includegraphics{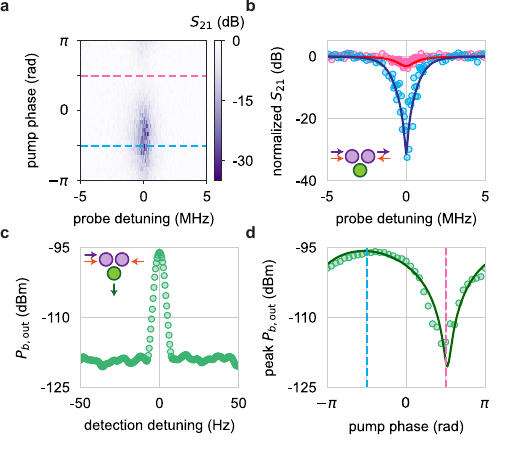} 
    \caption{\textbf{Characterization of isolation and circulation performance.} 
    \textbf{a,} Normalized $S_{21}$ as a function of probe frequency and relative phase between the two pumps.
    The data are normalized to background data acquired when $a_1$ and $a_2$ were far-detuned.
    Blue and red dashed lines mark relative phases where maximum isolation and transmission are achieved, respectively.
    \textbf{b,} $S_{21}$ traces at maximum isolating point (blue) and maximum passing point (red).
    \textbf{c,} Detection of (frequency-converted) circulation.
    Signal is injected from $a_1$ and output power ($P_{b,\text{out}}$) is measured using a spectrum analyzer.
    Detuning is given with respect to $\omega_b$ at the operation point.
    \textbf{d,} Peak output power as a function of relative pump phase. 
    Maximal and minimal emission occur at the same relative pump phases as isolation and pass in panel \textbf{a}.
    We note that while the shape and contrast of the power curve is predicted entirely by theory, the power offset of $\qty{-94}{dBm}$ has been estimated from the fridge wiring.
}
\label{fig:isolation and gyration}
\end{figure}

At the operation point, applying both conversion pumps simultaneously results in interference in the loop formed by $a_{1,2}$, $b$, and the transmission line.
Consequently, we expect phase-tuneable transmissions and reflections when injecting signals into a port, and monitoring emissions.
In principle, this would allow measuring the full S-matrix similar to the simulated one in \Cref{fig:overview}c.
Here, due to constraints in the wiring, we focus on $S_{21}$ and $S_{b1}$, which would directly correspond to using the device as an isolator or controllable absorber.
The phase-control of these quantities is confirmation that the directionality of the device can indeed be fully controlled.

To evaluate the isolation performance of the device, we have performed $S_{21}$ measurements using a VNA, while sweeping the relative phase between the two pumps. 
Pump powers are set such that the conversion rates are matched, $g_1 = g_2$. 
We note that both pumps are first tuned separately and then applied simultaneously with no further fine tuning. 
The small fourth order nonlinearity of the SNAIL modes minimizes interference between the two parametric processes and thus allows for a simple and robust tune-up procedure.
The normalized $S_{21}$ as a function of both pump phase and probe frequency, for $g_{1(2)}/2\pi = \qty{0.70}{MHz}$, is shown in \cref{fig:isolation and gyration}a. 
$S_{21}$ changes dramatically from low to high insertion loss at zero detuning, demonstrating clearly the parametrically tuneable directionality of the chiral coupler.
In \cref{fig:isolation and gyration}b, we show line cuts at the ``pass'' and ``isolation'' phase settings, revealing a narrow-band isolation of $\sim \qty{30}{\deci\bel}$ at the isolation setting.
We also observe a finite insertion loss of $\sim \qty{2}{\deci\bel}$ at the pass setting; this non-ideal behavior can be explained by internal decoherence, which may originate from flux noise (\cref{fig:ki vs phi}). 
We expect that this detrimental effect can be largely eliminated by a moderate adjustment of the circuit parameters (see next section for a detailed discussion). 
We emphasize that the isolation performance is predicted in quantitative agreement (blue and red lines in \cref{fig:isolation and gyration}b) using the \emph{same} parameters we have extracted from our calibration data shown in the previous section. 
We have repeated this measurement over a wider range of $g_{1(2)}$, with each measurement showing a similar level of agreement between data and theoretical prediction. 
The data shown in \cref{fig:isolation and gyration} were taken at the value for $g_{1(2)}$ the best peak isolation, in agreement with theory and limited by damping.
The model is presented in the \cref{app:S-matrix} and \cref{app:internal Q and flux noise}, and additional data are shown in \Cref{fig:isolation and gyration vs power}.

To directly evaluate circulation in the device, we have measured not only insertion loss between ports 1 and 2, but also the power emitted at port $b$, $P_{b,\text{out}}$, which is proportional to $S_{b1}$.
The data were taken using a spectrum analyzer, in the same phase sweep as the $S_{21}$ data.
In this measurement, the signal emitted at port $b$ occurs at the converted frequency $\omega_b = \omega_a + \omega_\text p$.
\Cref{fig:isolation and gyration}c shows detected power against detection frequency, taken at the `isolation' setting indicated in \cref{fig:isolation and gyration}a.
The fact that a prominent peak is visible at $\omega_b$ is an indicator that isolation in $S_{21}$ corresponds to transmission in $S_{b1}$, i.e., that circulation occurs.
Peak power as a function of phase is shown in \cref{fig:isolation and gyration}d.
The maximum emitted power coincides with the phase for the best isolation in $S_{21}$ (at $-90^{\circ}$), and similarly the minimum peak matches the best transmission (at $90^{\circ}$), providing strong evidence for circulation.
As with isolation, the circulation behavior is captured quantitatively, up to RF output line calibrations, by our model using the \emph{same} set of parameters. 
Data for additional values of $g_{1(2)}$ are presented in \cref{fig:isolation and gyration vs power}.


\section{Parameter-dependence of performance}
\label{sec:predictions}
The above described data and model confirm that our circuit is capable of achieving very good directional performance, and that it is highly predictable.
In the following, we use these insights as guide for predicting the performance of integrated devices that may be used for isolation and quantum signal routing.
Specifically, we are interested in how circuit parameters can be used to tailor insertion loss and isolation, as well as the efficiency with which the chiral coupler could be used to directionally emit and absorb quantum states when integrated with a qubit, as initially envisioned in \Cref{fig:overview}a.

\begin{figure}
\centering
\includegraphics{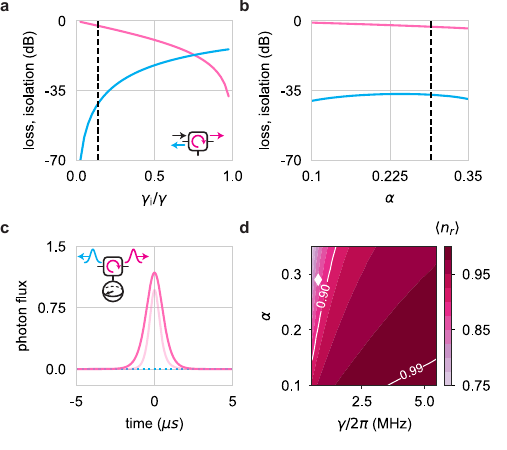}
\caption{\textbf{Performance predictions.}
    \textbf{a,} Insertion loss and isolation vs.\ damping and external coupling. 
    Parameters: $\gamma/2\pi=0.5$\,MHz, $\gamma_{b}/2\pi=0.5$\,MHz, $g/2\pi=0.5$\,MHz.
    \textbf{b,} Insertion loss and isolation as a function of SNAIL $\alpha$. 
    Dashed lines mark parameters for the device used here. 
    \textbf{c,} Photon flux (blue: left; red: right) when nominally emitting a single photon rightward as shaped wavepacket.
    Dark line: ideal device; bright line: parameters of the device used here.
    \textbf{d,} Fraction of photons emitted to the right as a function of external coupling rate and $\alpha$. 
    White marker: current device.
}
\label{fig:performance predictions}
\end{figure}

First, isolation and insertion loss are useful metrics to assess the utility of the chiral coupler as a circulator, whether stand-alone or integrated with quantum devices on a chip.
We have observed that the internal damping of the SNAIL modes causes a degradation of the device performance, resulting in finite isolation as well as non-zero insertion loss as shown in \cref{fig:isolation and gyration}b. 
The parameter governing this degradation is the ratio between internal and external damping, $\gamma_\text{i}/\gamma$. 
Here, we assume all the modes have the same external and internal damping rate.
In \cref{fig:performance predictions}a, we show the calculated isolation and insertion loss as a function of $\gamma_\text{i}/\gamma$, with device parameters similar to those used in the experiment presented above.
As expected, in the limit of no internal damping the device approaches ideal behavior, i.e., perfect isolation on resonance, and no insertion loss.

A smaller $\gamma_\text{i}/\gamma$ ratio can be achieved either by increasing the external coupling strength of the modes, or by reducing the internal damping rate.
Increasing the external coupling by moving the $a_{1,2}$ modes closer to the transmission line is straight-forward.
Regarding the internal damping, we have observed that measured linewidths increase significantly as frequencies are tuned away from the flux insensitive point.
This observation indicates that flux noise could be a limitation of coherence (see \cref{fig:ki vs phi}).
Assuming that flux noise is a dominating source of decoherence and that we cannot find an effective route to reduce it, we are left with the option to reduce flux sensitivity and tunability; this can be achieved by reducing the SNAIL parameter $\alpha$ \cite{frattini_3wave_2017} (see \cref{app:flux-noise}). 
In \cref{fig:performance predictions}b, we show the predicted isolation and insertion loss as a function of $\alpha$.
For the calculation we have assumed that the flux noise is the only source of internal broadening of the modes, and we model this as a loss rate that depends on frequency sensitivity, $\mathrm d\omega/\mathrm d\varphi_\text{ext}$.
We note that flux noise can typically be expected to result in dephasing rather than energy dissipation; an explicit distinction is beyond the scope of our current work.
It is worth noting that reducing $\alpha$ will also lead to a decrease in the third order non-linearity of the SNAIL, thus lowering the parametric coupling strength for the frequency conversion process; this can be compensated by stronger drives.
In summary, we predict that a modest change of design parameters $\gamma_\text{e}$ and $\alpha$ should be able to yield a dramatic improvement in both insertion loss and isolation.

Finally, we return to our originally envisioned use case for the device, namely to act as an interface for directionally controllable emission and absorption of arbitrary quantum states encoded in traveling wavepackets.
Based on our model above, we compute the predicted efficiency of on-demand, directional photon emission from $b$. 
We assume that $b$ is initialized with 1 photon, and by applying temporally shaped pumps with $g_{1(2)}(t)$ and appropriately chosen phase difference, the photon leaves as a shaped wavepacket in the desired direction.
For specifically chosen pump setting, the emitted photon flux as a function of time is shown in \cref{fig:performance predictions}c. 
While the directionality of emission is near-perfect using our experimentally established parameters, efficiency is suppressed due to internal loss and dephasing.
Similar to the case of insertion loss discussed above, however, we predict that high emission efficiency can be achieved by tailoring circuit parameters.
In \cref{fig:performance predictions}d we show the calculated emitted number of photons as a function of $\alpha$ and external coupling strength $\gamma$, where we assume the photons are emitted from an ideal $b$ mode without decoherence. 
Smaller $\alpha$ and larger $\gamma$ result in dramatically reduced photon loss in the device. 
With moderate adjustment of these parameters ($\gamma/2\pi \gtrsim 3.5$~MHz and $\alpha = 0.1$), we predict that directional photon emission and absorption with efficiencies of 99$\%$ is achievable.

\section{Conclusions and outlook}
\label{sec:conclusion}
We have demonstrated a versatile directional interface for integration with superconducting quantum circuits.
In this chiral coupler, parametrically controlled interference breaks time-reversal symmetry and realizes nonreciprocity with in-situ control.
The experimentally realized device displays a high degree of directionality, and we have verified its performance in both isolation and circulation measurements.
Our circuit design consists of a minimal number of modes, and can be described by a simple model.
With only few system parameters that can be calibrated independently, our model successfully captures the measured device performance quantitatively and without free fit parameters.
Based on this model, we predict that sub-percent inefficiency is within reach when using the chiral coupler as a qubit-integrated circulator or quantum signal router, without need for significant improvement in device quality.
Looking forward, directly integrating our chiral coupler with qubits on-chip will be a stepping stone for harnessing nonreciprocity for scalable quantum processors.
Integration into readout circuitry, the coupler may enable an isolator-free qubit readout scheme~\cite{chapman_widely_2017, abdo_active_2019, beck_wideband_2023, kwende_josephson_2023, navarathna_passive_2023}.
On the other hand, using the coupler as a quantum signal routing element, it can enable driven-dissipative remote entanglement~\cite{stannigel_drivendissipative_2012, pichler_quantum_2015, lingenfelter_exact_2023}, and the transfer of arbitrary quantum states in all-to-all connected quantum networks~\cite{axline_ondemand_2018, campagne-ibarcq_deterministic_2018, kurpiers_deterministic_2018, ai_multinode_2022}.

\section*{Acknowledgements}
We acknowledge funding from the NSF Quantum Leap Challenge Institute for Hybrid Quantum Architectures and Networks (Award 2016136) and from the IBM-Illinois Discovery Accelerator Institute.
Measurements were performed on a device fabricated by the Superconducting Qubits at Lincoln Laboratory (SQUILL) Foundry at MIT Lincoln Laboratory, with funding from the Laboratory for Physical Sciences (LPS) Qubit Collaboratory.
We thank R.~McDermott, S.~Chakram and B.~Du for advice on flux lines; 
S.~Mandal, A.~Baptista, S.~Cross, and S.~Rani for experimental assistance; 
and S.~Shankar and A.~Kou for critical reading of the manuscript.


\appendix

\section{Model}
\label{app:qo-model}

Here we give a detailed derivation of our model for the chiral coupler. 
We first treat the chiral coupler as a three-port device and derive the input-output relations for each port. 
Then the full S-matrix of the device is derived using these input-output relations.
We then discuss the waveguide-mediated coupling and how to cancel it with a bus-mediated coupling. 
Finally, we discuss the model for quantum state transfer in a cascaded system of two chiral emitters/absorbers implemented with our chiral coupler.
The symbols used in the following sections are summarized in \cref{tab:all-symbol}. 

\vfill\eject

\begin{table}[t!]
    \caption{\label{tab:all-symbol}
    Symbols used in the device model.
    }
    \begin{ruledtabular}
    \begin{tabular}{p{0.14\linewidth} | p{0.82\linewidth}}
    \textrm{Sym.}  & \textrm{Desc.} \\
    \colrule
    $\omega_0$ & frq.\ of $a_1$ and $a_2$ at operation point\\
    $\omega_b$ & frq.\ of $b$ at operation point\\
    $a_{1(2)}^{(\dagger)}$ & ladder operators for $a_{1(2)}$\\
    $b^{(\dagger)}$ & ladder operators for $b$\\
    $\omega_\text{p}$ & pump frq.\ for the conversion process\\
    $g_{1(2)}(t)$ & rate of conversion between $a_{1(2)}$ and $b$\\
    $\varphi_{1(2)}(t)$ & phase of conversion between $a_{1(2)}$ and $b$\\
    $d_\text{R(L)}^{(\dagger)}$ & ladder ops.\ for right (left) propagating modes in the common transmission line\\
    $d_b^{(\dagger)}$ & ladder ops.\ for the propagating mode in the semi-infinite transmission line on $b$'s port\\
    $\gamma_j$ & external coupling of $a_j$ to left and right propagating modes\\
    $\gamma$ & external coupling of both $a$ modes if they are equal\\
    $\gamma_{\text{i},j}$ & internal damping of mode $a_j$ \\
    $\gamma_{\text{i},a}$ & internal damping of both $a$ modes if equal\\
    $\gamma_b$ & external coupling of $b$ \\
    $\gamma_{i, b}$ & internal coupling of $b$ \\
    $x_j$ & position of $a_j$ along the common transmission line \\
    $v$ & speed of light in the transmission line \\
    \end{tabular}
    \end{ruledtabular}
\end{table}

\subsection{Input-output relations}
\label{app:input-output}
We first assume that there is no internal damping in the device.
We treat the chiral coupler as a three-port device, as shown in \cref{fig:overview}b,d.
The full Hamiltonian can be written as:
\begin{equation}
    H = H_{\text{sys}} + H_{\text{b}} + H_{\text{sb}},
    \label{Eq:H_full}
\end{equation}
$H_{\text{sys}} = H_\text{s} + H_\text{p} + H_\text{c}$ describes the system Hamiltonian for the chiral coupler.
Here, $H_\text{s}$ describes the $a$ and $b$ modes:
\begin{equation}
    H_{\text{s}} = \omega_{0} a_1^{\dagger} a_1 + \omega_b b^{\dagger} b + \omega_{0} a_2^{\dagger} a_2, 
    \label{Eq:H_s}
\end{equation}
$H_\text{p}$ describes the parametric frequency conversion process:
\begin{equation}
    H_{\text{p}} = g_{1}(t) e^{i (\omega_\text p t + \varphi_{1})} a_1^{\dagger} b + g_{2}(t) e^{i(\omega_\text p t + \varphi_{2})} a_2^{\dagger} b + \text{h.c.},
    \label{Eq:H_p}
\end{equation}
and $H_{\text{c}}$ describes the bus-mediated coupling between $a_1$ and $a_2$ mode:
\begin{equation}
    H_{\text{c}} = g_\text{c} (a_1^{\dagger} a_2 + a_1 a_2^{\dagger}).
    \label{Eq:H_c}
\end{equation}
For clarity we write $H_\text{c}$ as a separate term, rather than absorbing it into the system Hamiltonian.
It is used to cancel the waveguide-mediated coupling; its origin and how cancellation is achieved is discussed further below.

The second term on the left-hand side of \cref{Eq:H_full} is the bath Hamiltonian:
\begin{equation}
    H_\text{b} = \int d\omega~\omega (d_\text{R}^{\dagger}(\omega) d_\text{R}(\omega) + d_\text{L}^{\dagger}(\omega) d_\text{L}(\omega) +  d_b^{\dagger}(\omega) d_b(\omega)).
\end{equation}
The transmission line mode operators satisfy commutation relations $[d_j(\omega), d_{j'}^{\dagger}(\omega')] = \delta(\omega - \omega') \delta_{j, j'}$, where $j, j' = \text{L}, \text{R}, b$.
The system-bath interaction $H_{\text{sb}}$ is given by:
\begin{widetext}
\begin{equation}
    H_{\text{sb}} = i \frac{1}{\sqrt{2 \pi}} \int d\omega \left[ \sum_{j=1, 2} \sqrt{\gamma_j} (e^{i \omega \frac{x_j}{v}}  d_\text{L}^{\dagger}(\omega)  a_j + e^{-i\omega \frac{x_j}{v}} d_\text{R}^{\dagger}(\omega)  a_j ) + \sqrt{\gamma_b} d_b^\dagger (\omega)  b - \text{h.c.} \right].
    \label{Eq:system-bath Hamiltonian}
\end{equation}
\end{widetext}
Here we use the first Markov approximation, that is, the coupling strength $\gamma_j$ is independent of frequency. 

The input-output relation for the $b$ port can be obtained in the usual way~\cite{gardiner_input_1985}:
\begin{equation}
    d_b^{\text{out}}(t) - d_b^{\text{in}}(t) = \sqrt{\gamma_b} b(t),
    \label{Eq:db_input_output}
\end{equation}
and the dynamics for $b$ are described by:
\begin{equation}
    \Dot{b} = -i[b, H_{\text{sys}}] - \frac{\gamma_b}{2} b -\sqrt{\gamma_b} d_b^{\text{in}}(t).
    \label{Eq:b_dot}
\end{equation}

The equations of motion for mode $a_1$ and $a_2$ and input-output relation of the transmission line modes can be obtained in a similar way.
We first set $x_1 = 0$ and define $\tau = x_2/v$ for simplicity (without loss of generality).
The equation of motion for $d_k(\omega)$ ($k = \text{L}, \text{R}$) is given by:
\begin{align}
    \Dot{d_k} &= -i [d_k, H] \nonumber \\
    &= -i [d_k, H_\text{b}+H_{\text{sb}}] \nonumber \\
    &= -i \omega d_k + \sum_{j=1, 2} ( \sqrt{\frac{\gamma_j}{2\pi}} e^{(-1)^k i\omega \frac{x_j}{v}} a_j ).
    \label{Eq:d_k_dot_equation}
\end{align}
The solution is:
\begin{widetext}
\begin{equation}
    d_k(\omega, t) = e^{-i \omega (t - t_0)} d_k(\omega, t_0) + \sum_{j=1, 2} \left[\sqrt{\frac{\gamma_j}{2\pi}} \int_{t_0}^t dt' e^{-i\omega (t - t')} e^{ (-1)^k i \omega \frac{x_j}{v}} a_j(t') \right],
    \label{Eq:d_k_dot}
\end{equation}
\end{widetext}
where $d_k(\omega, t_0)$ is the initial value for $d_k$ at frequency $\omega$.
Similarly, the equation of motion for $a_j$ is given by:
\begin{widetext}
\begin{align}
    \Dot{a}_j &= -i [a_j, H] \nonumber \\
    &= -i [a_j, H_{\text{sys}}] - i [a_j, H_\text{b}+H_{\text{sb}}] \nonumber \\
    &= -i [a_j, H_{\text{sys}}] - \frac{1}{\sqrt{2\pi}} \int d\omega \sqrt{\gamma_j} ( e^{-i\omega \frac{x_j}{v}} d_\text{L}(\omega, t) +  e^{i\omega \frac{x_j}{v}} d_\text{R}(\omega, t)).
    \label{Eq:a_j_dot}
\end{align}
\end{widetext}
To solve the equation for $a_j$, we need the last two terms in \cref{Eq:a_j_dot}, which follow from \cref{Eq:d_k_dot}.
For $d_\text{L}$, for example, we obtain:
\begin{widetext}
\begin{align}
     &\frac{1}{\sqrt{2\pi}} \int d\omega \sqrt{\gamma_j} e^{-i \omega \frac{x_j}{v}} d_\text{L}(\omega, t) \nonumber \\
     = &\frac{1}{\sqrt{2\pi}} \int d\omega \sqrt{\gamma_j} e^{-i \omega \frac{x_j}{v}} ( e^{-i \omega (t - t_0)} d_\text{L}(\omega, t_0) + \sum_{j'=1, 2} \left[\sqrt{\frac{\gamma_{j'}}{2\pi}} \int_{t_0}^t dt' e^{-i\omega (t - t')} e^{i\omega \frac{x_{j'}}{v}} a_{j'}(t') \right] ) \nonumber \\
     = &\frac{1}{\sqrt{2\pi}} \int d\omega \sqrt{\gamma_j}  e^{-i \omega (t + \frac{x_j}{v} - t_0)} d_\text{L}(\omega, t_0)  + \frac{1}{2\pi} \int d\omega \int_{t_0}^t dt' \sum_{j'=1, 2} \left[ \sqrt{\gamma_j \gamma_{j'}} e^{-i \omega (t - t' + \frac{x_j}{v} - \frac{x_{j'}}{v})} a_{j'}(t') \right]. 
    \label{Eq:d_L_int0}
\end{align}
\end{widetext}
For simplicity, we set $x_1 < x_2$. 
We write the input field modes propagating in the $\text{L(R)}$ direction as:
\begin{equation}
    d_k^{\text{in}} = \frac{1}{\sqrt{2\pi}} \int d\omega e^{- i\omega (t - t_0)} d_k(\omega, t_0).
\end{equation}
Then \cref{Eq:d_L_int0} can be simplified to:
\begin{align}
&\frac{1}{\sqrt{2\pi}} \int d\omega \sqrt{\gamma_j} e^{-i \omega \frac{x_j}{v}} d_\text{L}(\omega, t) \nonumber \\ 
= &\sqrt{\gamma_j} d_\text{L}^{\text{in}}(t+\frac{x_j}{v}) + \frac{\gamma_j}{2}  a_j(t) \nonumber \\ 
+ &\sqrt{\gamma_j \gamma_{\Bar{j}}} a_{\Bar{j}}(t-\frac{|x_j - x_{\Bar{j}}|}{v}) \Theta(\Bar{j}-j),
\label{Eq:d_L_int}
\end{align}
where $\Bar{j} = 1$ if $j = 2$, and vice versa.
$\Theta(j-j')$ is the Heaviside step-function. 
We have used the $\delta$-function representation
\begin{equation}
 \int_{-\infty}^{\infty} d\omega e^{-i\omega (t-t')} = 2 \pi \delta(t-t')   
\end{equation}
in the above derivation. 
The integral for $d_\text{R}$ can be obtained in the same way:
\begin{align}
&\frac{1}{\sqrt{2\pi}} \int d\omega \sqrt{\gamma_j} e^{i \omega \frac{x_j}{v}} d_\text{R}(\omega, t) \nonumber \\
= &\sqrt{\gamma_j} d_\text{R}^{\text{in}}(t-\frac{x_j}{v}) + \frac{\gamma_j}{2}  a_j(t) \nonumber \\ 
&+ \sqrt{\gamma_j \gamma_{\Bar{j}}} a_{\Bar{j}}(t-\frac{|x_j - x_{\Bar{j}}|}{v})  \Theta(j-\Bar{j}).
\label{Eq:d_R_int}
\end{align}
We use the free-evolution approximation, $O(t-\frac{|x_j - x_{j'}|}{v}) = O(t) e^{i\omega \frac{|x_j - x_{j'}|}{v}}$. 
This approximation is valid when the delay time for photons from $a_1$ to $a_2$ is much shorter than the time scale of the evolution of the system.
The operator equation for $a_j$ can then be written as:
\begin{widetext}
\begin{equation}
    \Dot{a}_j = -i[a_j, H_{\text{sys}}] - \gamma_j a_j(t) - \sqrt{\gamma_j \gamma_{\Bar{j}}} e^{-i\omega_0 \frac{|x_j - x_{\Bar{j}}|}{v}}  a_{\Bar{j}} - \sqrt{\gamma_j} e^{-i \omega_0 \frac{x_j}{v} } d_\text{L}^{\text{in}} - \sqrt{\gamma_j} e^{i \omega_0 \frac{x_j}{v} } d_\text{R}^{\text{in}}.
    \label{Eq:a_dot_final}
\end{equation}
\end{widetext}
The term $\sqrt{\gamma_j \gamma_{\Bar{j}}} e^{i\omega_0 (t-\frac{|x_j - x_{\Bar{j}}|}{v})}  a_{\Bar{j}}$ corresponds to the waveguide-mediated coupling $H_j$.
In order to derive the input-output relation for the propagating mode in the transmission line, we note that the solution to \cref{Eq:d_k_dot_equation} can be written differently if we choose to integrate from a time $t_\text{f} > t$:
\begin{widetext}
\begin{equation}
    d_k(\omega, t) = e^{-i \omega (t - t_\text{f})} d_k(\omega, t_\text{f}) - \sum_{j=1, 2} \left[\sqrt{\frac{\gamma_j}{2\pi}} \int_{t_0}^{t_\text{f}} dt' e^{-i\omega (t - t')} e^{ (-1)^k i \omega \frac{x_j}{v}} a_j(t') \right].
\end{equation}
\end{widetext}
Consider the integral $\frac{1}{\sqrt{2\pi}} \int d\omega d_\text{L}(\omega, t)$:
\begin{widetext}
\begin{align}
    &\frac{1}{\sqrt{2\pi}} \int d\omega d_\text{L}(\omega, t) \nonumber \\
    = &\frac{1}{\sqrt{2\pi}} \int d\omega e^{-i\omega(t-t_0)} d_\text{L}(\omega, t_0) + \frac{1}{2\pi} \int d\omega \int_{t_0}^t dt' \sum_{j=1, 2} \left[ \sqrt{\gamma_j } e^{-i \omega (t - t'  - \frac{x_j}{v})} a_j(t') \right] \nonumber \\
    = &\frac{1}{\sqrt{2\pi}} \int d\omega e^{-i\omega(t-t_\text{f})} d_\text{L}(\omega, t_\text{f}) - \frac{1}{2\pi} \int d\omega \int_{t_0}^{t_\text{f}} dt' \sum_{j=1, 2} \left[ \sqrt{\gamma_j } e^{-i \omega (t - t'  - \frac{x_j}{v})} a_j(t') \right],  
\end{align}
\end{widetext}
and we write the output modes as:
\begin{equation}
    d_k^{\text{out}} = \frac{1}{\sqrt{2\pi}} \int d\omega e^{-i\omega(t-t_\text{f})} d_k(\omega, t_\text{f}).
\end{equation}
We then arrive at the input-output relation for the leftward propagating modes:
\begin{equation}
    d_\text{L}^{\text{out}} - d_\text{L}^{\text{in}} = \sum_{j=1, 2} \sqrt{\gamma_j} e^{i\omega_0 \frac{x_j}{v}}a_j(t).
    \label{Eq:dL_input_output}
\end{equation}
Similarly, we obtain for the rightward propagating modes:
\begin{equation}
    d_\text{R}^\text{out} - d_\text{R}^{in} = \sum_{j=1, 2} \sqrt{\gamma_j} e^{-i\omega_0 \frac{x_j}{v}}a_j(t).
    \label{Eq:dR_input_output}
\end{equation}
The spatial separation between $a_1$ and $a_2$ mode is $\lambda/4$, and we can set
$x_1 = 0, x_2 = \lambda/4$.
By applying this condition into \cref{Eq:db_input_output}, \cref{Eq:b_dot}, \cref{Eq:a_dot_final}, \cref{Eq:dL_input_output}, \cref{Eq:dR_input_output} and using the expression for $H_\text{s}$, $H_\text{p}$, $H_\text{c}$, we recover the input output relation and equation of motion discussed in the Model section of the main text.

\subsection{S-matrix}
\label{app:S-matrix}
The S-matrix of the chiral coupler can be obtained by solving the input-output relation and operator equations together in the frequency domain. 
We start from \cref{Eq:a equation}, \cref{Eq:b equation}, and \cref{Eq:c equation}.
Taking the Fourier transform on both sides, we obtain:
\begin{widetext}
\begin{align}
    -i\tilde{\omega}_0 a_1(\tilde{\omega}_0) = &-i\omega_0 a_1(\tilde{\omega}_0) -i g_{1} e^{i \varphi_{1}} b(\tilde{\omega}_0+\omega_\text p) - \gamma_1 a_1(\tilde{\omega}_0) \nonumber \\
    &- i \sqrt{\gamma_1 \gamma_2} a_2(\tilde{\omega}_0) - i g_\text{c} a_2(\tilde{\omega}_0) - \sqrt{\gamma_1} d_\text{L}^{\text{in}}(\tilde{\omega}_0) - \sqrt{\gamma_1}  d_\text{R}^{\text{in}}(\tilde{\omega}_0) \\
    \label{Eq:a1 equation}
    -i\tilde{\omega}_b b(\tilde{\omega}_b) = &-i\omega_b b(\tilde{\omega}_b) -i g_{1} e^{-i \varphi_{1}} a_1(\tilde{\omega}_b-\omega_\text{p}) -i g_{2} e^{-i\varphi_{2}} a_2(\tilde{\omega}_b - \omega_\text p) \nonumber \\
    &- \frac{\gamma_b}{2} b(\tilde{\omega}_b) - \sqrt{\gamma_b} b^{\text{in}}(\tilde{\omega}_b) \\
    -i\tilde{\omega}_0 a_2(\tilde{\omega}_0) = &-i\omega_0 a_2(\tilde{\omega}_0) -i g_{2} e^{i \varphi_{2}} b(\tilde{\omega}_0+\omega_\text p) - \gamma_2 a_2(\tilde{\omega}_0)  \nonumber \\
    &- i \sqrt{\gamma_1 \gamma_2}  a_1(\tilde{\omega}_0) - i g_\text{c} a_1(\tilde{\omega}_0) - \sqrt{\gamma_2} e^{-i\frac{\pi}{2}} d_\text{L}^{\text{in}}(\tilde{\omega}_0) - \sqrt{\gamma_2} e^{-i\frac{\pi}{2}} d_\text{R}^{\text{in}}(\tilde{\omega}_0),
\end{align}
\end{widetext}
where $\tilde{\omega}_0$ and $\tilde{\omega}_b$ are probe frequencies for the $a_{1(2)}$ and $b$ modes; these are frequencies close to the resonant frequencies $\omega_0$ and $\omega_b$, in practice.
The parametric conversion process requires the frequency matching condition $\omega_\text p = \omega_b - \omega_0 = \tilde{\omega}_b - \tilde{\omega}_0$; we define a frequency detuning $\delta = \tilde{\omega}_0 - \omega_0 = \omega_b - \tilde{\omega}_b$.
Then we obtain: 
\begin{widetext}
\begin{align}
    i\delta a_1 -i g_{1} e^{i \varphi_{1}} b - \gamma_1 a_1 - i \sqrt{\gamma_1 \gamma_2}  a_2 - i g_\text{c} a_2 - \sqrt{\gamma_1} d_\text{L}^{\text{in}} - \sqrt{\gamma_1}  d_\text{R}^{\text{in}} &= 0,  \\
    i\delta b -i g_{1} e^{-i \varphi_{1}} a_1 -i g_{2} e^{-i\varphi_{2}} a_2 - \frac{\gamma_b}{2} b - \sqrt{\gamma_b} b^{\text{in}} &= 0, \\
    i\delta a_2 -i g_{2} e^{i \varphi_{2}} b - \gamma_2 a_2 - i \sqrt{\gamma_1 \gamma_2}  a_1 - i g_\text{c} a_1 - \sqrt{\gamma_2} e^{-i\frac{\pi}{2}} d_\text{L}^{\text{in}} - \sqrt{\gamma_2} e^{-i\frac{\pi}{2}} d_\text{R}^{\text{in}} &= 0.
\end{align}
\end{widetext}
We omit the frequency dependent for the operators for simplicity, as now they are all measured at their corresponding probing frequencies.
These equations together with input output relations yield the S-matrix of the chiral coupler.
The S parameters are given by (here, for simplicity, let $\gamma_1 = \gamma_2 = \gamma_b = \gamma$, $g_1 = g_2 = g$, and $\gamma = -g_\text{c}$):

\begin{widetext}
\begin{equation}    
\begin{array}{ccc}
S_{11} = \frac{4 \frac{\gamma}{g} \cos{(\Delta \varphi)} }{A}    
, &
S_{12} = \frac{4i \frac{\gamma}{g}\sin{(\Delta \varphi)} -4 \frac{\delta}{g} + C}{A}, &
S_{13} = \frac{2(D+E)}{A}, \\
S_{21} = \frac{4i \frac{\gamma}{g}\sin{(\Delta \varphi)} -4 \frac{\delta}{g} + C }{A}, \label{Eq: S21 eq} &
S_{22} = \frac{4\frac{\gamma}{g} \cos{(\Delta \varphi)}}{A}, &
S_{23} = \frac{2(D-E)}{A}, \\
S_{31} = \frac{2 \frac{\gamma}{g}(i e^{-i\varphi_{1}} - e^{-i\varphi_{2}})}{B}, &
S_{32} = \frac{2 \frac{\gamma}{g}(i e^{-i\varphi_{1}} + e^{-i\varphi_{2}})}{B}, &
S_{33} = \frac{4 - \frac{(2\delta - i\gamma)(\delta + i\gamma)}{g^2}}{B}. 
\end{array}
\end{equation}  
\end{widetext}
where we have introduced $\Delta \varphi = \varphi_1 - \varphi_2$, $A = (-4 + \frac{(\delta + i \gamma)(2 \delta + i \gamma)}{g^2})(\frac{\delta + i \gamma}{g})$, $B = 4 - \frac{(2\delta + i \gamma)^2}{g^2}$, $C = \frac{(2\delta + i\gamma)(\delta^2 + \gamma^2)}{g^3}$, $D = e^{i\varphi_{1}} \frac{\gamma (-i \delta + \gamma)}{g^2}$, and $E = e^{i\varphi_{2}} \frac{\gamma (\delta + i \gamma)}{g^2}$. 
Each S parameter is obtained with other input signals set to 0 and the above result is used to generate the S-matrix in \cref{fig:overview}c with $\Delta \varphi = \pm \pi/2$.
For the more realistic case where $\gamma_1 \neq \gamma_2$, a similar solution can be obtained by choosing $g_c = -\sqrt{\gamma_1 \gamma_2}$.

\subsection{Cancellation of waveguide-mediated coupling}
\label{app:coupling cancellation}
As explained previously, there is a waveguide-mediated coupling between the $a_1$ and $a_2$ modes that needs to be cancelled. 
The coupling term arises from the second term on the right hand side of \cref{Eq:a_j_dot}, which is due to the interaction between $a_j$ and bath (the common transmission line) modes.
This term reads:
\begin{equation}
  \Dot{a}_j = \cdots  - \sqrt{\gamma_j \gamma_{\Bar{j}}} e^{-i\omega_0 \frac{|x_j - x_{\Bar{j}}|}{v}}  a_{\Bar{j}} + \cdots,
\end{equation}
which can be effectively written as 
\begin{equation}
    \Dot{a}_j = \cdots -i [ -i \sqrt{\gamma_j \gamma_{\Bar{j}}} e^{-i\omega_0 \frac{|x_j - x_{\Bar{j}}|}{v}} a_j^{\dagger} a_{\Bar{j}} + \text{h.c.}, a_j] + \cdots.
\end{equation}
Specifically for our case, where $|x_1 - x_2| = \lambda/4$ and assuming $\gamma_j = \gamma$, we have:
\begin{align}
    \Dot{a}_j &= \cdots -i [ -i \gamma e^{-i\frac{\pi}{2}} a_j^{\dagger} a_{\Bar{j}} + \text{h.c.}, a_j] + \cdots \nonumber \\
    &= \cdots -i [ - \gamma  a_j^{\dagger} a_{\Bar{j}} + \text{h.c.}, a_j] + \cdots.
\end{align}
Therefore this term acts as an effective interaction term between $a_1$ and $a_2$, which may be absorbed into the system Hamiltonian as an extra coupling term: $H_\text{wg} =  - \gamma ( a_1^{\dagger} a_{2} + a_1 a_{2}^{\dagger})$.
In the derivation of this interaction, we only assume a general system-bath interaction, as given in \cref{Eq:system-bath Hamiltonian}. 
As shown explicitly in \cref{fig: residual coupling}, this extra coupling is detrimental to the directionality of the chiral coupler, and needs to be eliminated. 

The cancellation is realized by exploiting a `quantum bus' coupling~\cite{majer_coupling_2007, chen_qubit_2014, lu_universal_2017, yan_tunable_2018} between the $a_1$ and $a_2$ modes through the $b$ mode. 
To understand the origin of this interaction, let us consider the three modes $a_1$, $a_2$ and $b$, with coupling between each other. 
The Hamiltonian can be written as:
\begin{align}
    H_\text{bus} &= \omega_0 a^{\dagger}_1 a_1 + \omega_0 a^{\dagger}_2 a_2 + \omega_b b^{\dagger} b \nonumber \\ 
    &+ (g_{12} a^{\dagger}_1 a_2 + g_b a^{\dagger}_1 b + g_b a^{\dagger}_2 b + \text{h.c.}),
\end{align}
where for simplicity we assume that the $a$ modes have the same frequency ($\omega_0$) and the same coupling ($g_b$) to the $b$ mode. 
We also assume a dispersive coupling regime, i.e., $g \ll \Delta$, where $\Delta = \omega_b - \omega_0$ is the frequency detuning between the modes.
We can then apply a Schrieffer-Wolff transformation.
The transfer matrix is given by:
\begin{equation}
    S = \frac{g_b}{\Delta} (a_1^{\dagger} b - a_1 b^{\dagger} +  a_2^{\dagger} b - a_2 b^{\dagger})
    \label{Eq:SW matrix}
\end{equation}
The Hamiltonian transforms as
\begin{equation}
    \Bar{H}_{\text{bus}} = e^{S} H_{\text{bus}} e^{-S},
    \label{Eq:Schrieffer-Wolff transformation}
\end{equation}
and we use the Baker-Campbell-Haussdorf formula:
\begin{equation}
    \Bar{H}_{\text{bus}} = H_{\text{bus}} + [S, H_{\text{bus}}] + \frac{1}{2} [S, [S, H_{\text{bus}}]] + \cdots.
    \label{Eq:SW expansion}
\end{equation}
Since $g \ll \Delta$, we can treat the mode couplings as a perturbation, and write the Hamiltonian as:
\begin{equation}
    H_{\text{bus}} = H^0_{\text{bus}} + V,
\end{equation}
where $H_{\text{bus}}^0 = \omega_0 a^{\dagger}_1 a_1 + \omega_0 a^{\dagger}_2 a_2 + \omega_b b^{\dagger} b $ and $V = (g_{12} a^{\dagger}_1 a_1 + g_b a^{\dagger}_1 b + g_b a^{\dagger}_2 b + \text{h.c.})$. 
Up to the first order in $V$, the transformed Hamiltonian in the dressed basis is now:
\begin{align}
    \Bar{H}_{\text{bus}} &= \omega'_0 a^{\dagger}_1 a_1 + \omega'_0 a^{\dagger}_2 a_2 + \omega'_b b^{\dagger} b \nonumber \\
    &+ \left [g_{12} + \frac{g^2_b}{\Delta} \right ] (a^{\dagger}_1 a_2 + a_1 a^{\dagger}_2 ), 
\end{align}
where $\omega'_0 = \omega_0 - g^2_b/\Delta$, and $\omega'_b = \omega_b + g^2_b/\Delta$ are the new mode frequencies in the dressed basis. 
Note that we have already worked in this dressed basis in the main text and derivations in previous sections, with the $'$ omitted.
This extra coupling between the two $a$ modes, $g^2_b/\Delta$ is generated because both $a_1$ and $a_2$ modes are coupled to the $b$ mode. 
Together with the original coupling between the two $a$ modes, we reach the ``cancellation coupling'', $g_\text{c}$, in the main text:
\begin{equation}
    g_\text{c} = g_{12} + \frac{g^2_b}{\Delta}.
    \label{Eq:cancellation coupling}
\end{equation}
Its frequency tunable nature allows us to reach the desired coupling strength by biasing $b$  to the right frequency. 
Equation~\cref{Eq:cancellation coupling} is used for fitting the cancellation ratio data shown in \cref{fig:basic calibration}c. 

\subsection{On-demand directional photon emission and absorption}

\begin{figure}[tb]
    \centering
    \includegraphics[scale = 1.0]{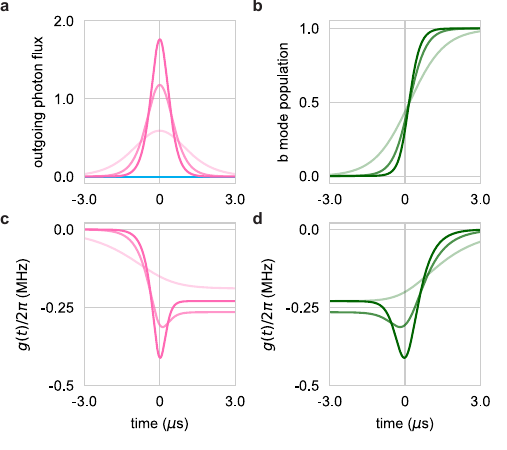}
    \caption{\textbf{Numerical solutions for on-demand photon emission and absorption.}
        \textbf{a,} Photon flux to the left (blue) and right (pink) as a function of time. 
        Colors from light to dark represent $\gamma_{\text{ph}} = 0.5 \gamma, 1.0\gamma, 1.5\gamma$ respectively. 
        \textbf{b,} Average photon number in $b$ mode as a function of time during the photon absorption process. Color from light to dark represents $\gamma_{\text{ph}} = 0.5 \gamma, 1.0\gamma, 1.5\gamma$ respectively. 
        \textbf{c, d,} Parametric pump amplitudes as a function of time for on-demand pitch and catch of the wave packets in \textbf{a}.
        Parameters used in the calculation: $\gamma/2\pi = 1$~MHz, $\gamma_b/2\pi = 0$~MHz, $\gamma_{\text{ph}} = 0.5 \gamma$, $1.0\gamma$, $1.5 \gamma$, $\varphi_{1} = \pi/2$, $\varphi_{2} = -\pi/2$.
    }
    \label{fig:Pitch and catch calculation}
\end{figure}

To investigate quantum signal routing, we consider the case of on-demand photon emission and absorption:
$b$ is initialized with one excitation, and we aim to release this excitation coherently as traveling photon, with directional control.
Vice versa, we want the ability to absorb traveling wavepackets into excitations in $b$.
In general, this requires a tunable coupling between the mode and its bath~\cite{cirac_quantum_1997}.
These tasks can be realized with the chiral coupler by parametrically controlling the couplings $g(t)$ between the $b$ mode and two $a$ modes. 

Here, we provide a semi-classical time-domain solution to this question by solving the input-output relations and equations of motion for the mode operators.
The target wave packet is chosen to have the form
\begin{equation}
    \phi(t) = \frac{\sqrt{\gamma_{\text{ph}}}}{2} \text{sech}\left(\frac{\gamma_{\text{ph}}}{2} t\right),
\end{equation}
where $1/\gamma_{\text{ph}}$ is the full width at half maximum for the wave packet in the time domain.
The form is chosen because it provides an analytic solution for the shape of the parametric pumps \cite{gheeraert_programmable_2020}; other shapes may be solved for numerically.
For the on-demand directional photon emission, we first initialize the $b$ mode in a coherent state with one photon on average, i.e., $\left<b^{\dagger} b\right> = 1$. 
In in the equations of motion and input-output relations, we set all inputs to 0, and solve with initial condition $a_1 = 0, a_2 = 0, b = 1$. 
The analytically obtained form of the magnitude of the parametric pump is given by:
\begin{equation}
    |g_{1(2)}(t)| = \frac{-\gamma_{\text{ph}} \sqrt{\gamma_{\text{ph}}} \tanh({\frac{\gamma_{\text{ph}}}{2}} t) + 2 \gamma \sqrt{\gamma_{\text{ph}}}} {-8 \sqrt{-\frac{\gamma_{\text{ph}}}{8} + \gamma \frac{(1-\tanh({\frac{\gamma_{\text{ph}}}{2}}t)) \cosh^2{(\frac{\gamma_{\text{ph}} }{2}}t)}{2} }}.
    \label{Eq:g(t)}
\end{equation}
The solutions for the pumps and emitted field as a function of time, for the case of a rightward emitted wavepacket, are shown in \cref{fig:Pitch and catch calculation}a,c. 
Directionality is achieved by controlling the relative phases between the two pumps.

On-demand photon absorption can be calculated in a similar way.
Because the incoming wave packet $\phi(t)$ is time-symmetric, the absorption process is the time-inverse of the emission. 
Absorption can thus be achieved by setting $g_{1(2)}(t) \rightarrow g_{1(2)}(-t)$. 
When solving the equations of motion, we set $d_\text{R}^{\text{in}}(t) = \phi(t)$, keep all other inputs 0, and initial conditions are $a_1 = 0, a_2 = 0, b = 0$. 
Example solutions for photon absorption are shown in \cref{fig:Pitch and catch calculation}b,d.

These calculations provide semi-classical evidence for directional emission and absorption capability of the chiral coupler.
To discuss its performance (e.g., transfer fidelity) when routing quantum signals, we present a quantum description below.

\subsection{Quantum state transfer}

We begin by deriving the quantum master equation for a single chiral coupler. 
Similar to \cref{Eq:a_j_dot}, we now consider the equation for an operator $O$:
\begin{equation}
    \Dot{O} = -i[O, H_{\text{sys}}] - i[O, H_\text{b} + H_{\text{sb}}].
\end{equation}
The second term on the RHS yields
\begin{widetext}
\begin{align}
    &~~-i[O, H_\text{b} + H_{\text{sb}}] \nonumber \\
    &= \frac{1}{\sqrt{2\pi}} \int d\omega  \sum_{j=1, 2}\sqrt{\gamma_j} (e^{i\omega \frac{x_j}{v}} d_\text{L}^{\dagger} [O, a_j] + e^{-i\omega \frac{x_j}{v}} d_\text{R}^{\dagger} [O, a_j] - e^{-i\omega \frac{x_j}{v}} [O, a_j^{\dagger}] d_\text{L}  - e^{i\omega \frac{x_j}{v}}  [O, a_j^{\dagger}] d_\text{R}) \nonumber \\
    &+\frac{1}{\sqrt{2\pi}} \int d\omega \sqrt{\gamma_b} (d_b^{\dagger} [O, b] - [O, b^{\dagger}] d_b) \nonumber \\
    &= \sum_{j=1, 2} \sqrt{\gamma_j} \left( (e^{i\omega_0 \frac{x_j}{v}} (d_\text{L}^{\text{in}})^{\dagger} + e^{-i\omega_0 \frac{x_j}{v}} (d_\text{R}^{\text{in}})^{\dagger} ) [O, a_j] - [O, a_j^{\dagger}] (e^{-i\omega_0 \frac{x_j}{v}} d_\text{L}^{\text{in}} + e^{i\omega_0 \frac{x_j}{v}} d_\text{R}^{\text{in}}) \right) \nonumber \\
    &+\sum_{j=1, 2} \gamma_j \left( a_j^{\dagger} [O, a_j] - [O, a_j^{\dagger}] a_j  \right) + \sum_{j=1, 2} \sqrt{\gamma_j \gamma_{\Bar{j}}} \left( e^{-i\omega_0 (t-\frac{|x_j - x_{\Bar{j}}|}{v})} a_{\Bar{j}}^{\dagger} [O, a_j] -  e^{i\omega_0 (t-\frac{|x_j - x_{\Dot{j}}|}{v})}  [O, a_{j}^{\dagger}] a_{\Bar{j}}\right) \nonumber \\
    &+\sqrt{\gamma_b} \left( (b^{\text{in}})^{\dagger} [O, b] - [O, b^{\dagger}] b^{\text{in}} \right) + \frac{\gamma_b}{2} \left(b^{\dagger} [O, b] - [O, b^{\dagger}] b \right),
\end{align}
\end{widetext}
where we used \cref{Eq:d_L_int} and \cref{Eq:d_R_int}, and a similar result for $b^\text{in}$ from input-output theory.
Now consider the expectation value of this operator $\left<O\right>$ in the Schrodinger picture.
We use the cyclic invariance of the trace operation and we also assume the initial state of the propagating mode to be the vacuum state, thus setting $b^{\text{in}}\rightarrow0$, $d_\text{L}^{\text{in}}\rightarrow0$, and $d_\text{R}^{\text{in}}\rightarrow0$:
\begin{widetext}
\begin{align}
    &\frac{d\left<O\right>}{dt} = \text{Tr} \left[\frac{dO}{dt} \rho_{\text{tot}} \right] = \text{Tr}\left[O \frac{d\rho_{\text{tot}}}{dt}\right] \nonumber \\
    &= -i\text{Tr} \biggl[O, [H_{\text{sys}}, \rho_{\text{tot}}]  +  \sum_{j=1, 2} 2\gamma_j [O (a_j \rho_{\text{tot}} a_j^{\dagger} - \frac{1}{2} \{ a_j^{\dagger} a_j, \rho_{\text{tot}}\}) ] + \gamma_b O (b \rho_{\text{tot}} b^{\dagger} - \frac{1}{2} \{ b^{\dagger} b, \rho_{\text{tot}}\}) \nonumber \\
    &+\sum_{j=1, 2} 2\sqrt{\gamma_j \gamma_{\Bar{j}}} O  \cos{(\omega_0 \frac{|x_j - x_{\Bar{j}}|}{v})}    \left(a_j \rho_{\text{tot}} a_{\Bar{j}}^{\dagger}- \frac{1}{2} (\rho_{\text{tot}} a_j^{\dagger} a_{\Bar{j}} + a_j^{\dagger} a_{\Bar{j}} \rho_{\text{tot}})\right) \nonumber \\
    &-i  \sqrt{\gamma_1 \gamma_2} O \sin{(\omega_0 \frac{|x_1 - x_2|}{v})}  [a_1^{\dagger} a_2 + a_1 a_2^{\dagger}, \rho_{\text{tot}}] \biggr].
\end{align}
\end{widetext}
Because the operator $O$ is an arbitrary local operator of the chiral coupler, its expectation value can be obtained independently of the partial trace of the propagating modes. 
Thus, the reduced master equation for the chiral coupler can be written as:
\begin{widetext}
\begin{equation}
    \frac{d \rho}{dt} = -i[H_{\text{sys}} + H_j', \rho] + \sum_{j=1, 2} (2 \gamma_j D(a_j) \rho + 2 \sqrt{\gamma_j \gamma_{\Bar{j}}} \cos{(\omega_0 \frac{|x_j-x_{\Bar{j}}|}{v})} D(a_j, a_{\Bar{j}}) \rho) + \gamma_b D(b) \rho,
\end{equation}
\end{widetext}
where $\rho = \text{Tr}_b(\rho_{\text{tot}})$ is the density matrix of the chiral coupler, $\text{Tr}_b(.)$ is the partial trace with respect to the propagating modes and $H_j' = \sqrt{\gamma_j \gamma_{\Bar{j}}} \sin{(\omega_0 \frac{|x_j - x_{\Bar{j}}|}{v})}  (a_1^{\dagger} a_2 + a_1 a_2^{\dagger})$ corresponds to the waveguide-mediated coupling between the $a$ modes.
When the physical separation between the $a$ modes is $\lambda/4$ and $\gamma_1 = \gamma_2 = \gamma$, the Hamiltonian will lead to the equations we used in the main text.

Next, we can derive from this the full quantum master equation of two chiral couplers that share the same transmission line.
The bath Hamiltonian for this case is:
\begin{align}
     H_\text{B} = \int d\omega~\omega ( &d_\text{R}^{\dagger}(\omega) d_\text{R}(\omega) + d_\text{L}^{\dagger}(\omega) d_\text{L}(\omega) \nonumber  \\ 
     + &d_{b_1}^{\dagger}(\omega) d_{b_1}(\omega) +  d_{b_2}^{\dagger}(\omega) d_{b_2}(\omega) ),
\end{align}
and the system-bath interaction Hamiltonian is:
\begin{widetext}
\begin{equation}
    H_{\text{SB}} = i \frac{1}{\sqrt{2 \pi}} \int d\omega \left[ \sum_{j=1}^4 \sqrt{\gamma_j} (e^{i \omega \frac{x_j}{v}}  d_\text{L}^{\dagger}(\omega)  a_j + e^{-i\omega \frac{x_j}{v}} d_\text{R}^{\dagger}(\omega)  a_j ) + \sum_{j=1}^2\sqrt{\gamma_{b_j}} d_{b_j}^\dagger (\omega)  b_j - \text{h.c.} \right].
\end{equation}
\end{widetext}
where $d^{(\dagger)}_{b_j}$ is the ladder operator for the propagating mode in the semi-
infinite transmission line on b’s port of the $j$th chiral coupler, and $\gamma_{b_j}$ is the external coupling of the corresponding $b$ mode. 
The input-output relations for the $b$ modes remain the same as the single coupler case. 
The solution for the propagating mode in the transmission line is given by:
\begin{widetext}
\begin{equation}
    d_k(\omega, t) = e^{-i \omega (t - t_0)} d_k(\omega, t_0) + \sum_{j=1}^4 \left[\sqrt{\frac{\gamma_j}{2\pi}} \int_{t_0}^t dt' e^{-i\omega (t - t')} e^{ (-1)^k i \omega \frac{x_j}{v}} a_j(t') \right].
\end{equation}  
\end{widetext}

Then, for an arbitrary local operator of the chiral coupler, we can obtain:
\begin{widetext}
\begin{align}
    &~~-i [O, H_\text{B} + H_{\text{SB}}] \nonumber \\
    &= \frac{1}{\sqrt{2\pi}} \int d\omega  \sum_{j=1}^4 \sqrt{\gamma_j} (e^{i\omega \frac{x_j}{v}} d_\text{L}^{\dagger} [O, a_j] + e^{-i\omega \frac{x_j}{v}} d_\text{R}^{\dagger} [O, a_j] - e^{-i\omega \frac{x_j}{v}} [O, a_j^{\dagger}] d_\text{L}  - e^{i\omega \frac{x_j}{v}}  [O, a_j^{\dagger}] d_\text{R}) \nonumber \\
    &~~+\frac{1}{\sqrt{2\pi}} \int d\omega \sum_{j=1}^2 \sqrt{\gamma_{b_j}} (d_{b_j}^{\dagger} [O, b_j] - [O, b_j^{\dagger}] d_{b_j}) \nonumber \\
    &= \sum_{j=1}^4 \sqrt{\gamma_j} \left( (e^{i\omega_0 \frac{x_j}{v}} (d_\text{L}^{\text{in}})^{\dagger} + e^{-i\omega_0 \frac{x_j}{v}} (d_\text{R}^{\text{in}})^{\dagger} ) [O, a_j] - [O, a_j^{\dagger}] (e^{-i\omega_0 \frac{x_j}{v}} d_\text{L}^{\text{in}} + e^{i\omega_0 \frac{x_j}{v}} d_\text{R}^{\text{in}}) \right) \nonumber \\
    &~~+\sum_{j=1}^4 \gamma_j \left( a_j^{\dagger} [O, a_j] - [O, a_j^{\dagger}] a_j  \right) + \sum_{j=1}^4 \sum_{k \neq j}^4 \sqrt{\gamma_j \gamma_k} \left( e^{-i\omega_0 (t-\frac{|x_j - x_k|}{v})} a_k^{\dagger} [O, a_j] -  e^{i\omega_0 (t-\frac{|x_j - x_k|}{v})}  [O, a_{j}^{\dagger}] a_k\right) \nonumber \\
    &~~+\sum_{j=1}^2 \left[\sqrt{\gamma_{b_j}} \left( (b_j^{in})^{\dagger} [O, b_j] - [O, b_j^{\dagger}] b_j^{in} \right) + \frac{\gamma_{b_j}}{2} \left(b_j^{\dagger} [O, b_j] - [O, b_j^{\dagger}] b_j \right) \right].
\end{align}
\end{widetext}

Then consider the same expectation value $\left<O\right>$ and set all the initial state of the propagating mode to be the vacuum state:
\begin{widetext}
\begin{align}
     &\frac{d\left<O\right>}{dt} = \text{Tr} \left[\frac{dO}{dt} \rho_{\text{tot}} \right] = \text{Tr}\left[O \frac{d\rho_{\text{tot}}}{dt}\right] \nonumber \\
    &= -i\text{Tr} \biggl[O, [H_{\text{sys}}, \rho_{\text{tot}}]  +  \sum_{j=1}^4 2\gamma_j [O (a_j \rho_{\text{tot}} a_j^{\dagger} - \frac{1}{2} \{ a_j^{\dagger} a_j, \rho_{\text{tot}}\}) ] + \sum_{j=1}^2 \gamma_{b_j} O (b_j \rho_{\text{tot}} b_j^{\dagger} - \frac{1}{2} \{ b_j^{\dagger} b_j, \rho_{\text{tot}}\}) \nonumber \\
    &~~+\sum_{j=1}^4 \sum_{k \neq j}^4 2\sqrt{\gamma_j \gamma_k} O  \cos{(\omega_0 \frac{|x_j - x_k|}{v})}    \left(a_j \rho_{\text{tot}} a_k^{\dagger}- \frac{1}{2} (\rho_{\text{tot}} a_j^{\dagger} a_k + a_j^{\dagger} a_k \rho_{\text{tot}})\right) \nonumber \\
    &~~-i \sum_{j=1}^4 \sum_{k > j}^4 \sqrt{\gamma_j \gamma_k} O \sin{(\omega_0 \frac{|x_j - x_k|}{v})}  [a_j^{\dagger} a_k + a_j a_k^{\dagger}, \rho_{\text{tot}}] \biggr].   
\end{align}
\end{widetext}

This reduces to a master equation for the two connected chiral couplers:
\begin{align}
    \frac{d \rho}{dt} = &-i[H_{\text{sys}} + H_J', \rho] + \sum_{j=1}^4 2 \gamma_j D(a_j) \rho + \sum_{j=1}^2 \gamma_{b_j} D(b_{j}) \rho \nonumber \\ 
    &+ \sum_{j=1}^4 \sum_{k \neq j}^4 2 \sqrt{\gamma_j \gamma_k} \cos{(\omega_0 \frac{|x_j-x_k|}{v})} D(a_j, a_k) \rho,
    \label{Eq:two chiral coupler master eq}
\end{align}

where $H_J' = \sum_{j=1}^4 \sum_{k > j}^4 \sqrt{\gamma_j \gamma_k}  \sin{(\omega_0 \frac{|x_j - x_k|}{v})} (a_j^{\dagger} a_k + a_j a_k^{\dagger}) $ is the waveguide-mediated coupling between the $a$ modes.

\begin{figure}
\centering
\includegraphics{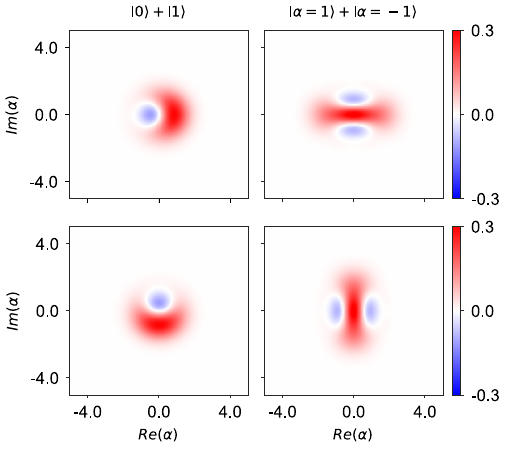}
\caption{\textbf{Quantum state transfer simulations.}
    The Wigner functions of the state sent (top row) via a first chiral coupler and received (bottom row) by a second chiral coupler are shown for two different states.
    Parameters used in the calculation are: $\gamma/2\pi = 5$~MHz, $\gamma^e_b/2\pi = 0$~MHz, $\gamma_{\text{ph}} = 0.1\gamma^a_e$.
}
\label{fig:quantum state transfer ideal}
\end{figure}

Using the master equation \cref{Eq:two chiral coupler master eq} we can now compute numerically state transfer fidelities.
To illustrate this here, we consider an ideal situation (no experimental imperfections), and two different types of initial states: Fock state superpositions, and cat states (i.e., coherent state superpositions). 
At the $t=0$, the $b$ mode in the first chiral coupler is initialized with the chosen initial state.
The parametric pumps for both chiral coupler are then activated, with a phase setting such that the first device emits the state to the right, while the second one absorbs radiation coming from left. 
The amplitude of the pumps are set according to \cref{Eq:g(t)}, with $t \rightarrow t$ and $t \rightarrow -t$ for emission and absorption processes respectively. 
The calculated Wigner functions for the states sent and received by the two chiral couplers are shown in \cref{fig:quantum state transfer ideal}. 
As the model only requires linear modes with parametric interactions, the chiral coupler is not limited within the basis of $\ket{0}$ and $\ket{1}$, but allows us to work with different encoding schemes, which shows its potential to route complex quantum states throughout a network. 

\section{Experimental setup}
\label{app:Exp setup}

The circuit was fabricated by the Superconducting Qubits at Lincoln Laboratory (SQUILL) Foundry at MIT Lincoln Laboratory.
The designed SNAIL junction parameters are: 
$L_j = 3.0$\,nH ($L_j$ value for one junction on the three-junction arm) and $\alpha = 0.29$ for $a_1$ and $a_2$; $L_j = 2.5~$nH and $\alpha = 0.29$ for $b$.

\begin{figure*}[p]
\centering
\includegraphics{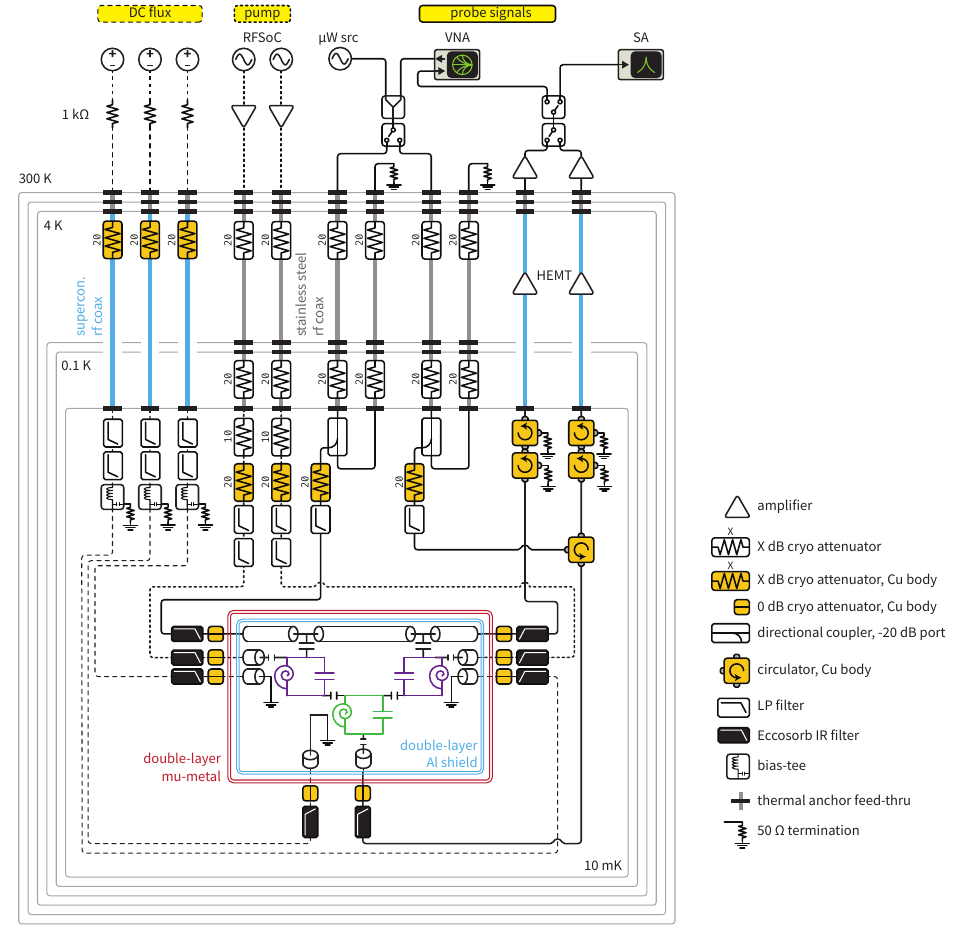}
\caption{\textbf{Experimental setup.}
    Most of the wiring and filtering follows typical best practices \cite{krinner_engineering_2019}.
    Relevant additional specifics are as follows:
    \textbf{Low-pass (LP) filters:} probe lines: 12\,GHz (\emph{K\&L 5L250-10200});
    pump lines: additional 2.8\,GHz LP filters (\emph{Mini-circuits ZLSS-A2R8G-S+});
    flux lines: two low pass filters (\emph{Mini-circuits VLFX-80} and \emph{Mini-circuits SLP 1.9}) and a bias tee (\emph{Mini-circuits ZFBT-4R2GW+}).
    \textbf{Output line configuration:} two cascaded double-stage isolators (\emph{Low Noise Factory LNF-ISCIC4\_12A}), HEMT amplifier (\emph{Low Noise Factory LNF-LNC4\_8C}); additional circulator (\emph{Low Noise Factory}) for reflectometry on mode $b$; room temperature low-noise amplifier (\emph{Low Noise Factory LNF-LNR4\_14C}).
    \textbf{Room temperature electronics:} 
    DC source for flux bias: \emph{Yokogawa GS200}; VNA: \emph{Keysight P9374A}; 
    RF source used in gyration measurements: \emph{SignalCore SC5511A};
    Spectrum analyzer (SA) used in gyration measurements: \emph{Keysight N9030B}.
    Pump tones were synthesized with a \emph{Xilinx} RFSoC board, programmed using the open-source QICK software \cite{stefanazzi_qick_2022}.
}
\label{fig:fridge setup}
\end{figure*}

Measurements were performed using an Oxford Triton 500 dilution refrigerator, at a base temperature of 10 mK. 
The experimental setup is shown in \cref{fig:fridge setup}; most of the wiring follows common best practices in circuit QED experiments \cite{krinner_engineering_2019}.

\section{Performance analysis and imperfections}
\label{sec:performance}
We have already established the model for the chiral coupler in the ideal case. 
We now go one step further and discuss performance limits imposed by non-ideal parameters.
We first discuss impact of internal damping rate, and how the performance can be improved by using a SNAIL with smaller $\alpha$. 
We then discuss the effect of imperfect cancellation of waveguide-mediated coupling.

\subsection{Effect of damping}
\label{app:internal Q and flux noise}

\begin{figure}
\centering
\includegraphics{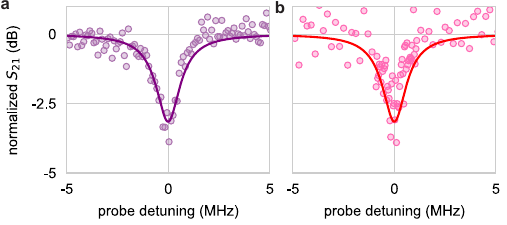}
\caption{
    \textbf{Insertion loss with and without pumps.}
    Normalized $S_{21}$ measured in two different cases at the best waveguide-mediated-coupling cancellation point.
    \textbf{a,} Both parametric pumps are off (the top data shown in \cref{fig:basic calibration}b). 
    \textbf{b,} Parametric pumps are on with a relative pump phase set to $\pi/2$ (the pink data shown in \cref{fig:isolation and gyration}b).
}
\label{fig:insertion loss with and without pumps}
\end{figure}

\begin{figure}
\centering
\includegraphics{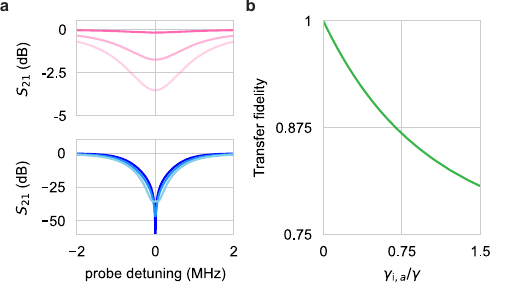}
    \caption{\textbf{Calculated impact of internal damping on the performance of the chiral coupler.}
    \textbf{a,} Numerical solution for $S_{21}$ at passing (top) and isolation (bottom) settings with finite internal damping rate; for simplicity we set $\gamma_{\text{i},1}=\gamma_{\text{i},2}\equiv\gamma_{\text{i},a}$.
    Passing (pink) and blocking (blue) predictions are shown as a function of probe frequency detuning at different $\gamma_{\text{i},a}$ vs $\gamma$ ratios. 
    The color from dark to light represent $\gamma_{\text{i},a}/\gamma = 0.01, 0.1, 0.2$ respectively. 
    \textbf{b,} The fidelity of transfering $(\ket{0}+\ket{1})/\sqrt{2}$ from one chiral coupler to another as a function of the ratio between $\gamma_{\text{i},a}$ and $\gamma$.
  }
\label{fig:insertion loss and isolation}
\end{figure}

Including the effect from finite internal damping, the relevant equations of motion become:
\begin{align}
    \dot{a}_1 = &-i\omega_0 a_1 -i g_{1} e^{i (\omega_\text{p} t + \varphi_{1})} b - \frac{(2\gamma  + \gamma_{\text{i},1})}{2} a_1\nonumber \\
    &- i \gamma a_2 - i g_\text{c} a_2 - \sqrt{\gamma}  d_\text{L}^{\text{in}} - \sqrt{\gamma}  d_\text{R}^{\text{in}}, \label{Eq:a equation2} \\
    \dot{b} = &-i\omega_b b -i g_{1} e^{-i (\omega_\text{p} t + \varphi_{1})} a_1  \nonumber \\ 
    &-i g_{2} e^{-i(\omega_\text{p} t +  \varphi_{2})} a_2 - \frac{(\gamma_b + \gamma_{\text{i},b})}{2} b - \sqrt{\gamma_b} b^{\text{in}}, \label{Eq:b equation2} \\
    \dot{a}_2 = &-i\omega_0 a_2 -i g_{2} e^{i(\omega_\text{p} t +  \varphi_{2})} b - \frac{(2 \gamma + \gamma_{\text{i},2})}{2} a_2 \nonumber \\ 
    &- i \gamma a_1 - i g_\text{c} a_1 - \sqrt{\gamma} e^{-i \frac{\pi}{2}} d_\text{L}^{\text{in}} - \sqrt{\gamma} e^{i \frac{\pi}{2}} d_\text{R}^{\text{in}} \label{Eq:c equation2},
\end{align}
where $\gamma_{\text{i},1}, \gamma_{\text{i},2}, \gamma_{\text{i},b}$ are the internal damping rates of $a_1, a_2, b$, respectively. 
To discriminate between the effect of finite internal $Q$ and detrimental effects from pumping we have measured $S_{21}$ under different conditions (\cref{fig:insertion loss with and without pumps}). 
We observe similar $S_{21}$, indicating that the parametric pumps do not introduce excess damping and the device insertion loss is a result of internal damping.

To further understand the behavior of the chiral coupler with a finite internal Q, we calculate the $S_{21}$ by solving this set of equations together with the input-output relations.
Predictions for $S_{21}$ with finite internal damping rates are shown in \cref{fig:insertion loss and isolation}a. 
The result shown in \cref{fig:performance predictions}a is obtained by finding the minimum of $S_{21}$ over a larger range of the $\gamma_{\text{i},a}/\gamma$ ratio.
The increase of the internal damping rate of the SNAIL mode thus clearly results in a decrease of both insertion loss and isolation of the chiral coupler.

This internal loss also impacts state transfer fidelity in a cascaded system.
We consider an example where the initial state $(\ket{0}+\ket{1})/\sqrt{2}$ is transferred from the first chiral coupler to the second, and the process is simulated using the full quantum model developed in the previous section.
The transfer fidelity as a function of $\gamma_{\text{i},a}/\gamma$ is shown in \cref{fig:insertion loss and isolation}b. 

\subsection{Flux noise}
\label{app:flux-noise}

\begin{figure}
\centering
\includegraphics[scale = 1.0]{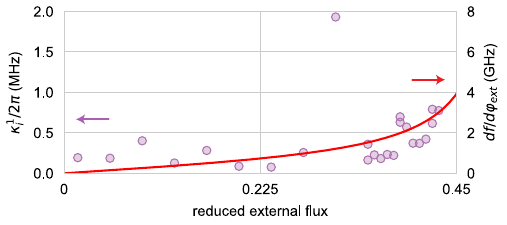}
    \caption{\textbf{SNAIL linewidth and flux sensitivity.}
    (Left axis, dots) Measured linewidth of $a_1$ as a function of external flux. (Right axis, line) Calculated flux sensitivity of the mode frequency as a function of external flux. 
  }
\label{fig:ki vs phi}
\end{figure}

\begin{figure}
\centering
\includegraphics[scale = 1.0]{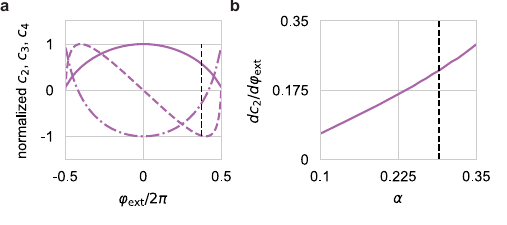}
    \caption{\textbf{SNAIL properties.}
    \textbf{a,} The SNAIL coefficients as a function of external flux. All the coefficients are normalized to their maximum values. 
    The solid line, dashed line, and dash-dot line represents the $c_2$, $c_3$, and $c_4$ values, respectively. 
    Ideally, the SNAIL is operated at the point where the fourth order nonlinearity is suppressed. 
    The black dashed line represents the operation point for the data shown in the main text.
    \textbf{b,} The slope of $c_2$ at the fourth order cancellation point as a function of SNAIL $\alpha$ value. 
    The black dashed line represents the parameter of our current device. 
  }
\label{fig:SNAIL calculation}
\end{figure}

Lowering damping of the internal modes of the chiral coupler would immediately result in improved performance.
As shown in \cref{fig:ki vs phi}, the measured linewidth of mode $a_1$ displays a trend that qualitatively matches its flux sensitvity.
Because there is an obvious difference ($\sim$ factor of 3) in the SNAIL linewidth between the flux insensitive point and the operation point, we believe that flux noise is a major reason for damping or dephasing, and thus of lowered performance.
One promising approach to reduce the impact of the flux noise is to use a SNAIL with a smaller $\alpha$. 
We note that energy loss and dephasing could not be distinguished in our network analyzer measurements, as both manifest as line broadening in the frequency domain.
An improved noise model could be established, for example, using ring-down measurements.

The SNAIL is designed to be a dipole circuit element with third order nonlinearity and minimal fourth order nonlinearity~\cite{frattini_3wave_2017}. 
A typical SNAIL is a superconducting loop of 3 Josephson junctions and a single smaller junction shunted by a large capacitor. 
The whole loop is threaded with magnetic flux $\Phi_{\text{ext}}$. 
The Hamiltonian for the SNAIL is:
\begin{equation}
    H_\text{S} = 4 E_\text C n^2 -\alpha E_\text J \text{cos}(\varphi) - 3 E_\text J \text{cos}(\frac{\varphi_{\text{ext}} - \varphi}{3}),
    \label{Eq:SNAIL H}
\end{equation}
where $E_\text C$ is the charging energy of the SNAIL mode, $n$ is the charge operator, $E_\text J$ is the Josephson energy is of the small junction and $\alpha$ is the ratio between Josephson energy between the large and small junctions, $\varphi$ is the phase over the small junction and $\varphi_{\text{ext}}$ is the reduced external flux. 
We an expand the inductive part $H^\text L_\text{S}$ of this Hamiltonian near the energy minimum point ($\varphi_\text{min}$):
\begin{align}
    H^\text{L}_{\text{S}} = \frac{E_J}{2} (&c_2 (\varphi - \varphi_{\text{min}})^2 + c_3 (\varphi - \varphi_{\text{min}})^3 \nonumber \\
    + &c_4 (\varphi - \varphi_{\text{min}})^4 + \cdots),
    \label{Eq:SNAIL expansion}
\end{align}
where the coefficients for each order are functions of $\alpha$ and $\varphi_{\text{ext}}$, $c_i = c_i(\alpha, \varphi_\text{{ext}})$. 
As an example, normalized SNAIL coefficients as a function of external flux calulated with the design values of $a_1$ are shown in \cref{fig:SNAIL calculation}a.

To estimate the SNAIL mode dephasing rate due to the flux noise, we assume the chiral coupler is operated at fourth order nonlinearity cancellation point.
Given the coefficients above, the SNAIL mode can then be quantized, with the resonant frequency $\omega_\text{S} = \sqrt{8 E_C (c_2 E_j)}$.
The dephasing rate of the SNAIL $\Gamma^\phi_S$ is a function of the flux noise power spectrum, $S_{\text{flux}}(\omega)$:
\begin{equation}
    \Gamma^\phi_\text{S} \propto \frac{d \omega_\text{S}}{d \varphi_{\text{ext}}} S_{\text{flux}}(\omega).
\end{equation}
Here we assume the flux noise is the same across the frequency tuning range. 
Then, the dephasing rate is set by the frequency sensitivity to the external flux, which is proportional to the derivative of SNAIL coefficient $c_2$ with respect to the external flux. 
This derivative is shown in \cref{fig:SNAIL calculation}b as a function of $\alpha$. 
In \cref{fig:performance predictions}d of the main text, the impact of $\alpha$ is estimated by setting the internal damping rate proportional to the value at $\alpha=0.29$ according to the calculated derivative value. 
We note that by choosing a SNAIL with $\alpha = 0.1$, we predict an improvement of a factor of 3 compared to our current device. 

\subsection{Effect of Kerr nonlinearity}
In order to match the $\lambda/4$ separation condition, it is difficult to operate the SNAIL exactly at the Kerr free point.
The black dashed line in \cref{fig:SNAIL calculation}a shows the operation flux value for the $a_{1, 2}$ mode, where the residual Kerr nonlinearity is $25\%$ of the maximum value.
However, the use of SNAIL largely suppresses this fourth order nonlinearity, we do not observe an obvious pump induced Kerr shift when operating the device.
We do note that the issue of matching the frequency for $\lambda/4$ separation to the Kerr free point does not put a fundamental limitation on the design and can be optimized with a more precise circuit parameter control in the fabrication process.

\subsection{Effect of pump leakage}
\begin{figure}
\centering
\includegraphics{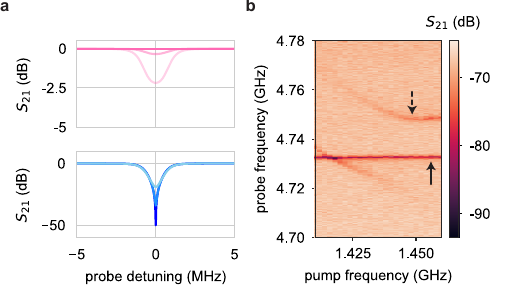}
    \caption{\textbf{Impact of pump leakage.}
    \textbf{a,} Numerical solution for $S_{21}$ at passing (top) and isolation (bottom) settings with different pump leakage ratio; for simplicity we set $\gamma_{\text{i},1}=\gamma_{\text{i},2}\equiv\gamma_{\text{i},a}$.
    Passing (pink) and blocking (blue) predictions are shown as a function of probe frequency detuning at different $\gamma_{\text{i},a}$ vs $\gamma$ ratios. 
    The color from dark to light represent $\epsilon = 0.0, 0.2, 0.5$ respectively. 
    \textbf{b,} Measured $S_{21}$ as a function of pump frequency and probe frequency.
    Both $a_1$ and $a_2$ are biased at the same frequency and the parametric pump for $a_1 \leftrightarrow b$ is applied to the system with a changing frequency.
    The dashed and solid arrow indicate the $a_1$ and $a_2$ mode respectively.
    }
    \label{fig:pump leakage}
\end{figure}

Pump leakage can be modeled as an extra mode conversion term in the equation of operators.
For example, including a pump leakage term in (\cref{Eq:a1 equation}) results in:
\begin{align}
    &-i\tilde{\omega}_0 a_1(\tilde{\omega}_0) \nonumber \\
    &= -i\omega_0 a_1(\tilde{\omega}_0) -i (g_{1} e^{i \varphi_{1}} + \epsilon g_{2} e^{i \varphi_{2}}) b(\tilde{\omega}_0+\omega_\text p) \nonumber \\
    &- \gamma_1 a_1(\tilde{\omega}_0) - i \sqrt{\gamma_1 \gamma_2} a_2(\tilde{\omega}_0) - i g_\text{c} a_2(\tilde{\omega}_0) \nonumber \\
    &- \sqrt{\gamma_1} d_\text{L}^{\text{in}}(\tilde{\omega}_0) - \sqrt{\gamma_1}  d_\text{R}^{\text{in}}(\tilde{\omega}_0),
\end{align}
where the term $\epsilon g_{2} e^{i \varphi_{2}} b(\tilde{\omega}_0+\omega_\text p)$ represents the effect of $a_2$'s pump on the $a_1$ mode. 
$\epsilon$ is the pump leakage ratio between the two pump lines. 
The pump leakage from $a_1$ to $a_2$ can be modeled similarly.
In \cref{fig:pump leakage}a, we show the predictions for $S_{21}$ with different pump leakage ratios. 
The increase of the pump leakage ratio clearly results in a decrease of both insertion loss and isolation of the chiral coupler.

To evaluate the pump leakage in our experiment, we have first biased both $a_1$ and $a_2$ to the same frequency and then applied the pump inducing mode conversion $a_1 \leftrightarrow b$.
The measured $S_{21}$ as a function of the pump frequency is shown in \cref{fig:pump leakage}b.
We can observe the pump induced mode splitting pattern on the $a_1$ mode while the $a_2$ mode remains unchanged under the pump.
This observation suggests that there is no noticeable pump leakage.

\subsection{Effect of imperfect waveguide-mediated-coupling cancellation}
\label{app:imperfect cancellation}

\begin{figure}
\includegraphics{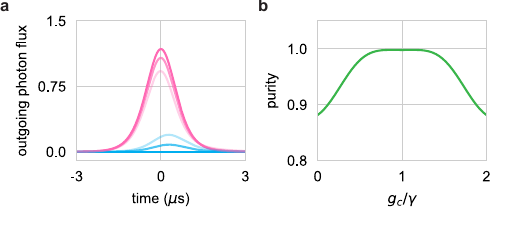}
    \caption{\textbf{Effect of residual waveguide-mediated coupling on directionality.}
    (a) Left (blue) and right (magenta) emitted photon flux as a function of time. 
    Color from light to dark represent $|g_\text{c}/\gamma| = 1.0, 0.8, 0.5$ respectively. 
    (b) Purity of the received state when transferring state $(\ket{0}+\ket{1})/\sqrt{2}$ as a function of $g_\text{c}/\gamma$.
  }
\label{fig: residual coupling}
\end{figure}

Another important factor that affects the performance of the chiral coupler is the imperfect cancellation of the waveguide-mediated coupling, i.e., $|g_\text{c}| \neq |\gamma|$.
The effect of this case be seen, for example, in imperfect directional emission of shaped wavepackets.
In \cref{fig: residual coupling}a we show the calculated outgoing photon flux as a function of time after $b$ is initialized with one photon, and drives are applied to emit the state.
The `pitched' photon is partially emitted into the wrong direction as the coupling cancellation deviates from the ideal case.
This leads to decoherence in a network: 
The purity of the state received by a second party  (when attempting to transfer $(\ket{0}+\ket{1})/\sqrt{2}$) as function of the ratio $g_\text{c}/\gamma$ is shown in \cref{fig: residual coupling}b.

\begin{figure}
\includegraphics[scale = 1.0]{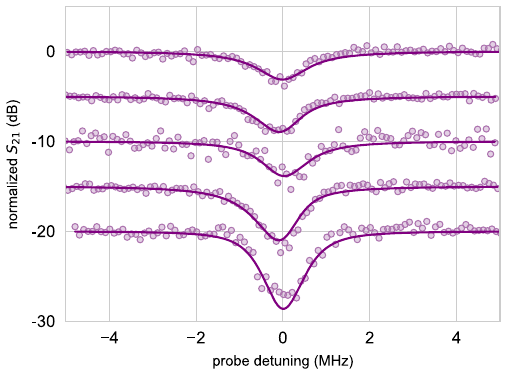}
    \caption{\textbf{Coupling cancellation measurement.}
    Note that we offset the traces by a 5 dB for visual clarity.
    The cancellation ratio $g_\text{c}/\gamma = 1.04, 0.72, 1.28, 0.55, 1.60$ from top to bottom correspondingly. 
    The degree of waveguide-mediated coupling cancellation is obtained by measuring $S_{21}$ at the operation point without pumping. 
    The dip of the curve reflects the degree of cancellation.
    The cancellation ratio is obtained by fitting the data to the model given in \cref{Eq: S21 eq nonideal}. 
  }
\label{fig:coupling cancellation}
\end{figure}

\subsection{Calibrating waveguide-mediate coupling cancellation}

The degree of the cancellation is inferred by measuring the $S_{21}$ trace when both $a$ modes are biased at operation point without applying pumps. 
Ideally, unit transmission is achieved at the cancellation point, and a dip will emerge as we deviate from it.
The depth of the dip reflects the degree of the cancellation. 
In practice, we still see an finite dip near the perfect cancellation point, due to the internal damping rate of the SNAIL mode.
With finite $\gamma_{\text{i},k}$ and $g_\text{c} \neq \gamma$,  $S_{21}$ at $g=0$ becomes
\begin{align}
    S_{21} = &[4 g^2_c + 8 g_\text{c} \gamma_\text{e}  -  4 \delta^2 + \gamma_{\text{i}, 1} \gamma_{\text{i}, 2} + 2 i \delta (\gamma_{\text{i}, 1} + \gamma_{\text{i}, 2})] / \nonumber  \\
    &[8 \gamma^2_e + 4 g^2_c + 8 g_\text{c} \gamma_\text{e} -  4 \delta^2 + \gamma_{\text{i}, 1} \gamma_{\text{i}, 2}  \nonumber \\
    &+ 2 i \delta (\gamma_{\text{i}, 1} + \gamma_{\text{i}, 2}) + 2 \gamma_\text{e} (4 i \delta + \gamma_{\text{i}, 1} + \gamma_{\text{i}, 2})].
    \label{Eq: S21 eq nonideal}
\end{align}
It is easy to verify that with $g_\text{c} = -\gamma_\text{e} = -\gamma$, $\gamma_{\text{i}, 1} = \gamma_{\text{i}, 2} = 0$, this turns back into the ideal case \cref{Eq: S21 eq}. 

In order to locate the cancellation point, we fit the data to the model given in \cref{Eq: S21 eq nonideal}.
We show $S_{21}$ measured at different cancellation points in \cref{fig:coupling cancellation}. 
The coupling cancellation ratio shown in \cref{fig:basic calibration}c in the main text is obtained from this fit. 
 
\subsection{Isolation and circulation performance}

\begin{figure}
\centering
\includegraphics[scale = 1.0]{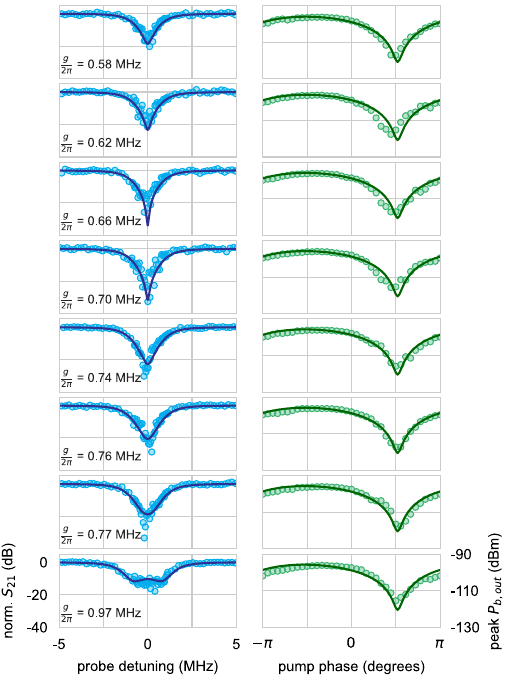}
\caption{
    \textbf{Isolation and gyration at different pump powers.}
    (left) Normalized $S_{21}$ data  as a function of probe frequency 
    (right) Peak output power at port $b$ as a function of probe frequency.
    A -94 dBm of power offset estimated from fridge wiring is added to the theory curve for the isolation measurement.
    Note that the isolation performance is not optimal for both small and large $g$ values; this is explained by the fact ideal nonreciprocity requires matching coherent and dissipative dynamics of the modes~\cite{clerk_introduction_2022}.
    }
\label{fig:isolation and gyration vs power}
\end{figure}

By taking into account all parameters and models discussed above, our theory shows good quantitative agreement with measured isolation and circulation data across a wider range of pump powers. 
In \cref{fig:isolation and gyration vs power}, we show both isolation and circulation data obtained at different pump powers. 
The theory curves are obtained with parameters independently calibrated as discussed in the main text. 
$g$ was obtained by extrapolation of the data shown in \cref{fig:basic calibration}e.
We stress that a single set of circuit parameters (\cref{tab:params} in the main text) was used for predicting all data in \cref{fig:isolation and gyration vs power}.

\newpage

\bibliography{references}
\end{document}